%% file: paper.tex
\documentclass[sigplan,nonacm,screen]{acmart}
\pdfoutput=1

\setcopyright{acmcopyright}
\copyrightyear{2018}
\acmYear{2018}
\acmDOI{XXXXXXX.XXXXXXX}

\acmConference[Conference acronym 'XX]{Make sure to enter the correct
  conference title from your rights confirmation emai}{June 03--05,
  2018}{Woodstock, NY}
\acmPrice{15.00}
\acmISBN{978-1-4503-XXXX-X/18/06}

\include{mcorelisting}

\usepackage{algorithm}
\usepackage[noend]{algpseudocode}
\usepackage{ebproof}

\usepackage{standalone}
\usepackage{tikz}
\usepackage{wrapfig}
\usepackage{cleveref}
\usepackage{subcaption}

\usetikzlibrary{patterns}

\input{commands}

\newcommand{\david}[1]{}

\usepackage{booktabs}   
\usepackage{subcaption} 

\algblockdefx[Match]{Match}{MatchEnd}
  [2]{\textbf{match} #1 \textbf{with} #2 \textbf{then}}
  [1]{\textbf{else} #1}
\algcblockdefx[Match]{Match}{MatchEls}{MatchEnd}
  [2]{\textbf{else match} #1 \textbf{with} #2 \textbf{then}}
  [1]{\textbf{else} #1}

\makeatletter
\newenvironment{btHighlight}[1][]
{\begingroup\tikzset{bt@Highlight@par/.style={#1}}\begin{lrbox}{\@tempboxa}}
{\end{lrbox}\bt@HL@box[bt@Highlight@par]{\@tempboxa}\endgroup}

\newcommand\btHL[1][]{%
  \begin{btHighlight}[#1]\bgroup\aftergroup\bt@HL@endenv%
}
\def\bt@HL@endenv{%
  \end{btHighlight}%
  \egroup
}
\newcommand{\bt@HL@box}[2][]{%
  \tikz[#1]{%
    \pgfpathrectangle{\pgfpoint{1pt}{0pt}}{\pgfpoint{\wd #2}{\ht #2}}%
    \pgfusepath{use as bounding box}%
    \node[anchor=base west, fill=orange!30,outer sep=0pt,inner xsep=1pt, inner ysep=0pt, rounded corners=3pt, minimum height=\ht\strutbox+1pt,#1]{\raisebox{1pt}{\strut}\strut\usebox{#2}};
  }%
}
\makeatother
\lstdefinestyle{MCore}{
  language={MCore},
  escapeinside={--}{\^^M},
  moredelim=**[is][{\btHL}]{@}{@},
}

\AtBeginDocument{\DeclareCaptionSubType{lstlisting}}
\crefname{sublstlisting}{listing}{listings}
\Crefname{sublstlisting}{Listing}{Listings}

\begin{document}

\title{Expression Acceleration: Seamless Parallelization of Typed High-Level Languages}


\author{Lars Hummelgren}
\affiliation{
  \department{Digital Futures and EECS}              
  \institution{KTH Royal Institute of Technology}            
  \city{Stockholm}
  \country{Sweden}                    
}
\email{larshum@kth.se}          

\author{John Wikman}
\affiliation{
  \department{Digital Futures and EECS}              
  \institution{KTH Royal Institute of Technology}            
  \city{Stockholm}
  \country{Sweden}                    
}
\email{jwikman@kth.se}          

\author{Oscar Eriksson}
\affiliation{
  \department{Digital Futures and EECS}              
  \institution{KTH Royal Institute of Technology}            
  \city{Stockholm}
  \country{Sweden}                    
}
\email{oerikss@kth.se}          

\author{Philipp Haller}
\orcid{0000-0002-2659-5271}             
\affiliation{
  \department{Digital Futures and EECS}              
  \institution{KTH Royal Institute of Technology}            
  \city{Stockholm}
  \country{Sweden}                    
}
\email{phaller@kth.se}          

\author{David Broman}
\orcid{0000-0001-8457-4105}             
\affiliation{
  \department{Digital Futures and EECS}              
  \institution{KTH Royal Institute of Technology}            
  \city{Stockholm}
  \country{Sweden}                    
}
\email{dbro@kth.se}          


\begin{abstract}
Efficient parallelization of algorithms on general-purpose GPUs is essential in many areas today. However, it is a non-trivial task for software engineers to utilize GPUs to improve the performance of high-level programs in general. Although many domain-specific approaches are available for GPU acceleration, it is difficult to accelerate existing high-level programs without rewriting parts of the programs using low-level GPU code. We present a compiler implementation using an alternative approach called \emph{expression acceleration}. This approach marks expressions for acceleration, and the compiler automatically infers which dependent code needs to be accelerated. We design and implement a compiler supporting expression acceleration for a statically typed functional language and evaluate its applicability and performance.
\end{abstract}

\ccsdesc[500]{Software and its engineering~Parallel programming languages}
\ccsdesc[300]{Software and its engineering~Source code generation}

\keywords{GPUs, acceleration, parallelization}  

\maketitle

\section{Introduction}
\label{sec:introduction}


Parallel computing on GPUs has been tremendously successful in many areas, including 3D graphical rendering, machine learning, and blockchains. However, efficient implementations on general-purpose GPUs require deep knowledge of low-level languages and frameworks, such as CUDA and OpenCL. Moreover, modern software systems are often developed in high-level languages, with closures, garbage collection, and user-defined recursive data types. Parallelizing subsets of such systems---if even possible---is typically very time-consuming. In particular, parts of the code need to be manually translated to low-level GPU code, including interactions between the low-level parallel code and the high-level program and low-level data marshaling.

There are many approaches to solving this semantic gap between low-level GPU implementations and high-level languages. For instance, specialized high-level languages and libraries enable efficient compilation of high-level parallel constructs~\cite{leissa2018anydsl,henriksen2017futhark,team2016theano,hagedorn2020achieving,steuwer2017lift}. Other works use a domain-specific language approach to achieve efficiency~\cite{kelley2013halide,sujeeth2014delite,svensson2008obsidian,chakravarty2011accelerating}. Such approaches typically emphasize efficient compilation of \emph{the whole} domain-specific language or library, and not the interaction with other high-level languages. On the other hand, for more machine-oriented languages, such as C and C++, there are approaches where code parts can be marked to be executed on either CPUs or on GPUs~\cite{openacc,openmp}. Fewer attempts have been made to \emph{partially} annotate high-level languages to accelerate existing sequential code automatically.

In contrast to previous work on imperative, dynamically-typed languages~\cite{lam2015numba,catanzaro2011copperhead,tillet2019triton,zhou2024appy,frostig2018compiling}, where functions are explicitly marked as being executed on the GPU, we introduce a compiler implementing the concept of \emph{expression acceleration} in a statically typed language setting. Specifically, the key idea is that any expression within a high-level language can be marked as accelerated. The compiler validates that the expression is supported and automatically resolves dependencies of accelerated expressions and inserts marshaling code, thus making the acceleration seamless to the end-user.


The rest of the paper is structured as follows. Section~\ref{sec:expression-acceleration} presents the idea of expression acceleration and describes the recommended compiler workflow. We provide an overview of the compiler pipeline (Section~\ref{sec:compilation-pipeline}) and discuss key analyses of the compiler (Section~\ref{sec:analysis-and-extraction}). We show that expression acceleration is practically usable and that it significantly improves performance (Section~\ref{sec:evaluation}). We discuss related work in Section~\ref{sec:related-work} and conclude in Section~\ref{sec:conclusions}.



\section{Expression Acceleration}
\label{sec:expression-acceleration}

In this section, we first present the key idea of expression acceleration, followed by a short outline of a practical approach to programming with expression acceleration. We implement expression acceleration in a language extended from the intermediate high-level functional language of the Miking framework~\cite{broman2019vision}. The extensions include parallel keywords used to support expression acceleration. The compiler implementation is available as open-source\footnote{\url{https://github.com/miking-lang/miking}}.

\subsection{Key Idea}
\label{subsec:key-idea}

Consider the program of Listing~\ref{lst:mot-ex} using expression acceleration. The program reads input (line~\ref{lst:mot-ex:read}), runs an \emph{accelerated expression} on lines~\ref{lst:mot-ex:cuda-acc1}-\ref{lst:mot-ex:cuda-acc2}, executes a sequential expression on line~\ref{lst:mot-ex:conv} before running another accelerated expression on line~\ref{lst:mot-ex:fut-acc}, and finally writes the result to a file (line~\ref{lst:mot-ex:write}).

The \mcoreinline{accelerate} keyword (lines~\ref{lst:mot-ex:cuda-acc1}-\ref{lst:mot-ex:cuda-acc2} and line~\ref{lst:mot-ex:fut-acc} in Listing~\ref{lst:mot-ex}) takes an accelerated expression as an argument, indicating that it should execute in parallel. Our compiler validates that this is supported, and automatically translates parallelism within accelerated expressions to execute on the GPU. The compiler has two separate backends; one is a functional backend that outputs Futhark~\cite{henriksen2017futhark} code and uses the Futhark compiler to produce GPU code (used on line~\ref{lst:mot-ex:fut-acc}), and the other is an imperative backend that produces CUDA code directly (used on lines~\ref{lst:mot-ex:cuda-acc1}-\ref{lst:mot-ex:cuda-acc2}).

The accelerated expression can have free variables. Consider the use of \mcoreinline{accelerate} on line~\ref{lst:mot-ex:fut-acc}, which refers to three free variables. The function \mcoreinline{computeSum} has to be included in the code we run in the accelerated context, and the values of \mcoreinline{t3} and \mcoreinline{n} have to be copied to the accelerated context. We outline the steps required to support this in Section~\ref{sec:analysis-and-extraction}.



Our extensions to the sequential source language consist of three categories of parallel constructs:
\begin{itemize}
\item The \mcoreinline{accelerate} keyword.
\item Functional parallel keywords, such as \mcoreinline{reduce} (line~\ref{lst:mot-ex:cond}).
\item Imperative parallel keywords, including \mcoreinline{loop} (line~\ref{lst:mot-ex:cuda-acc2}).
\end{itemize}

\begin{lstlisting}[
  style=MCore,
  caption={Example program making use of expression acceleration.},
  label={lst:mot-ex},
  float,
  floatplacement=t!,
  firstline=23,
  lastline=35
]
match readInput () with (t1, t2, n) in --\label{lst:mot-ex:read}
accelerate ( --\label{lst:mot-ex:cuda-acc1}
  let axpy = lam c. lam i.
    let x = tensorGet t1 [i] in
    let y = tensorGet t2 [i] in
    tensorSet t1 [i] (addi x (muli c y)) in
  loop n (lam i. axpy 2 i)); --\label{lst:mot-ex:parloop} \label{lst:mot-ex:cuda-acc2}
let t3 = create n (lam i. tensorGet t1 [i]) in --\label{lst:mot-ex:conv}
let computeSum = lam s. lam n. --\label{lst:mot-ex:sum1}
  if geqi n 100 then reduce addi 0 s --\label{lst:mot-ex:cond}
  else foldl addi 0 s in --\label{lst:mot-ex:sum2}
let sum = accelerate (computeSum t3 n) in --\label{lst:mot-ex:fut-acc}
writeOutput sum --\label{lst:mot-ex:write}
\end{lstlisting}

\subsection{Recommended Workflow}
\label{subsec:recommended-workflow}

\begin{figure}[t!]
\centering
\includegraphics[width=0.4\textwidth]{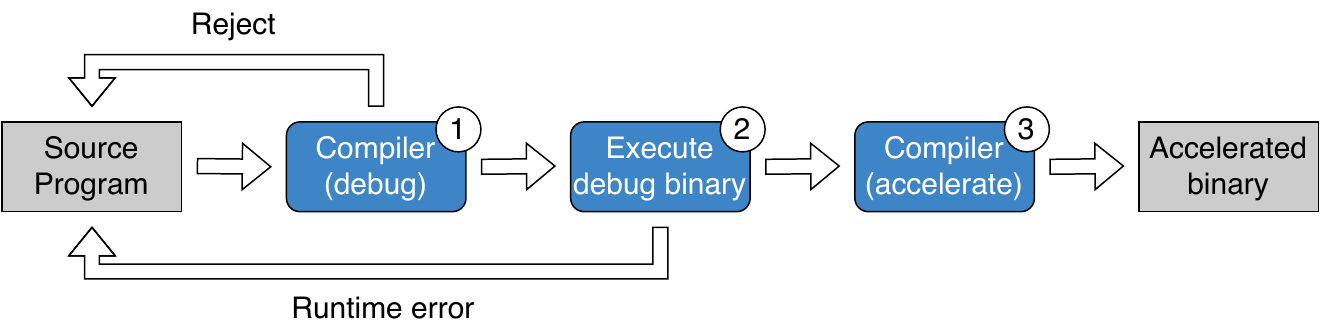}
\caption{The recommended workflow when using acceleration, from a source program on the left-hand side to an accelerated binary on the right-hand side. Gray boxes are artifacts and blue boxes are processes.}
\label{fig:workflow}
\end{figure}

Figure~\ref{fig:workflow} illustrates our proposed workflow which starts from a source program and ends with an accelerated binary. First, we compile the program in \emph{debug mode} (1 in the figure). The compiler validates the accelerated expressions but produces sequential CPU code with runtime checks for details that we cannot verify statically. The user executes the debug binary (2) to ensure no runtime errors occur. If the compilation or execution fails, the user must rewrite the source program. This workflow ensures informative error messages, which is challenging for code running on the GPU.

If the debug executable produces no errors, we compile the program in \emph{accelerate mode} (3), where parallel keywords in accelerated expressions execute on the GPU. We omit the line from this step to the source program because compilation errors are caught in (1). For efficiency reasons, we omit the runtime checks in the accelerated binary. This workflow enables early and direct feedback and produces a high-performance executable without unnecessary runtime overheads in the accelerated code.



\section{Compilation Pipeline}
\label{sec:compilation-pipeline}

\begin{figure*}[t!]
\centering
\includegraphics[width=0.85\textwidth]{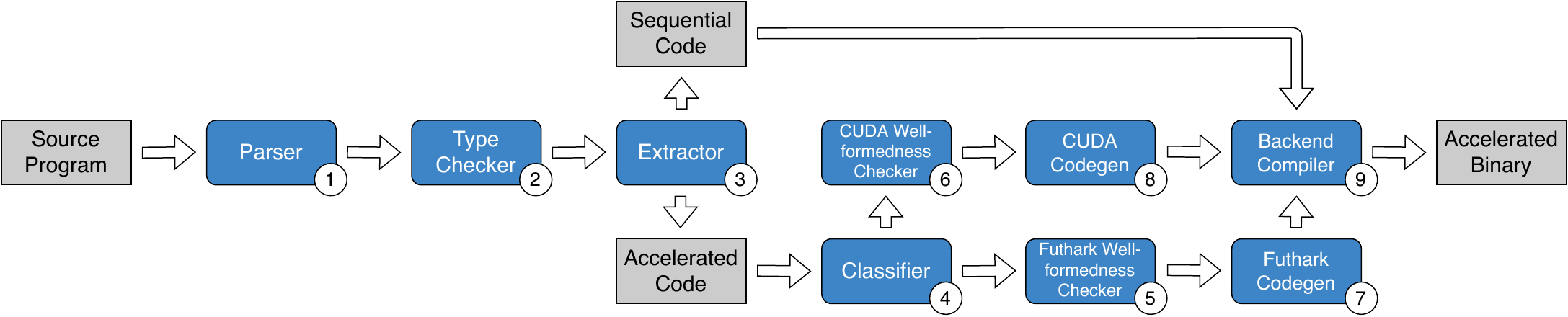}
\caption{Overview of the pipeline of the accelerate compiler. The gray rectangles represent artifacts, and the blue rounded rectangles represent compiler passes. Note that we omit most intermediate artifacts for brevity. We associate integers with each compiler pass to indicate the order in which it takes place.}
\label{fig:detailed-overview}
\end{figure*}

Figure~\ref{fig:detailed-overview} depicts the accelerate compiler pipeline, starting from a source program on the left and ending with an accelerated binary on the right. First, the compiler performs parsing and type-checking (1 and 2 in Figure~\ref{fig:detailed-overview}). This step is followed by \emph{accelerate extraction} in the extractor (3). In this step, we extract the accelerated expressions and the functions they depend on into a separate program (Section~\ref{subsec:extraction}).

Following the extraction, the program consists of two intermediate artifacts. The first part is the sequential code, which represents the parts of the program that execute sequentially. The second part is the accelerated code. This code is classified (Section~\ref{subsec:classification}) into backend-specific programs using a classifier (4). Our compiler supports two accelerate backends. The first emits Futhark~\cite{henriksen2017futhark} code, a pure functional array language. We use the Futhark compiler to produce efficient CUDA code. This backend supports a purely functional subset of the source language, operating primarily on immutable sequences. Futhark does not support imperative programs or operations with side effects. For this reason, we define another backend that directly emits CUDA code. This backend supports an imperative subset of the source language, primarily operating on mutable tensors.

The compiler must ensure these programs are supported by the respective accelerate backend. Thus, we define a set of well-formedness rules (Section~\ref{subsec:well-formedness}) for each backend that the compiler verifies statically (5 \& 6). Given that the code is well-formed, each backend generates GPU code (7 \& 8). Finally, the backend compiler (9) glues together the sequential code with the backend-specific programs by generating marshaling code based on the types of accelerated expressions (Section~\ref{subsec:marshaling}). The resulting binary, which we refer to as the \emph{accelerated binary} (right side of Figure~\ref{fig:detailed-overview}), runs the sequential parts on the CPU while accelerated expressions are offloaded to the GPU.

\section{Analysis and Extraction}
\label{sec:analysis-and-extraction}

In this section, we present important analyses in the compilation pipeline. We focus on the key step in supporting expression acceleration---the extraction (Section~\ref{subsec:extraction}). We also discuss the high-level ideas of classification (Section~\ref{subsec:classification}), data marshaling (Section~\ref{subsec:marshaling}), and well-formedness (Section~\ref{subsec:well-formedness}). For more details, see the technical paper (we will add a reference if this paper is accepted).

\subsection{Extraction}
\label{subsec:extraction}

We need to extract the parts of a program marked for acceleration into a separate AST. To make acceleration convenient, we allow an expression $\accelerate{e}$ to contain free variables. A naive approach to the extraction is to include all bound functions in the accelerated code. However, this includes more code than necessary, leading to increased compilation times. Also, it prevents sequential code from using features unsupported by the target accelerate backend. Therefore, our extraction includes only functions used in an accelerated expression. This enables sequential code to use all features available in the source language. It is also efficient as it avoids duplicating functions used in multiple accelerated expressions.

Extraction of accelerated expressions takes place in two steps. First, we rewrite accelerated bindings and apply lambda lifting to the source program. The second and key step is to emit the parts of the program used in accelerated expressions. In the extraction, we assume the names used in binding expressions (e.g., let-bindings) are unique. This can be achieved through a renaming pass. We refer to the unique name of a variable \mcoreinline{x} as an \emph{identifier}.

As an illustration of the extraction, consider the programs of Listing~\ref{lst:lamlift}. The input program, which contains an accelerated expression, is shown in Listing~\ref{lst:lamlift1}. In the first step of the accelerate extraction, we rewrite expressions \mcoreinline{accelerate e} as an \emph{accelerated binding} with identifier \mcoreinline{a}, such that \mcoreinline{let a = e in a}. We apply lambda lifting~\cite{johnsson1985lambda} modified to treat accelerated bindings \mcoreinline{a} as functions. The lambda lifting captures free variables of non-function type in $e$ and adds them as parameters of \mcoreinline{a}. Lambda lifting also lifts nested bindings to the top of the program, which is required by targets that do not support nested functions (e.g., CUDA).

Listing~\ref{lst:lamlift2} shows the program after applying the first step. Note that the accelerated expression is replaced with an accelerated binding \mcoreinline{a} and that the nested function \mcoreinline{f} is lifted outside of \mcoreinline{a}. Further, observe that the definitions of \mcoreinline{s} and \mcoreinline{c} are not included in the accelerated code. Instead, they are captured as arguments of \mcoreinline{a} by the lambda lifting (line~\ref{lst:lamlift2:acc1}), and passed to the function in the application on line~\ref{lst:lamlift2:sum}. This approach allows us to compute values in the sequential code, without the restrictions imposed on the accelerated code (e.g., the \mcoreinline{readSequence} function may perform I/O, which is not supported in accelerated code) and then pass them to the accelerated code.

\renewcommand{\figurename}{Listing}
\begin{figure*}[t!]
\centering
\begin{minipage}{.45\textwidth}
\centering
\subcaption{Input program making use of acceleration. This program is the input to the accelerate extraction.}
\label{lst:lamlift1}
\begin{lstlisting}[
  style=MCore,
  firstline=13,
  lastline=20
]
let g = lam x. addi x 1 in --\label{lst:lamlift1:g}
let s = readSequence () in
let c = randIntU 10 20 in
let sum = accelerate ( --\label{lst:lamlift1:acc1}
  let f = lam x. muli c (g x) in --\label{lst:lamlift1:f}
  let s2 = map f s in
  reduce addi 0 s2) in --\label{lst:lamlift1:acc2}
printSum sum
\end{lstlisting}
\end{minipage}\hfill
\begin{minipage}{.45\textwidth}
\centering
\subcaption{Program of Listing~\ref{lst:lamlift1} after applying the first step of the accelerate extraction. The accelerated code produced by extraction consists of the highlighted parts of the program.}
\label{lst:lamlift2}
\begin{lstlisting}[
  style=MCore,
  firstline=13,
  lastline=21
]
@let g = lam x. addi x 1 in@ --\label{lst:lamlift2:g}
let s = readSequence () in --\label{lst:lamlift2:s}
let c = randIntU 10 20 in --\label{lst:lamlift2:c}
@let f = lam c. lam x. muli c (g x) in@ --\label{lst:lamlift2:f}
@let a = lam s. lam c.@ --\label{lst:lamlift2:acc1}
  @let s2 = map (f c) s in@ --\label{lst:lamlift2:f-free}
  @reduce addi 0 s2 in@ --\label{lst:lamlift2:acc2}
let sum = a s c in --\label{lst:lamlift2:sum}
printSum sum --\label{lst:lamlift2:print}
\end{lstlisting}
\end{minipage}
\caption{Example program before and after applying the accelerate extraction.}
\label{lst:lamlift}
\end{figure*}
\renewcommand{\figurename}{Figure}

In the second step, we extract the parts of the program used within accelerated expressions into a separate program. The extracted program for the example consists of the highlighted parts of Listing~\ref{lst:lamlift2}. To arrive at this result, we go through the program in a bottom-up fashion, starting from the accelerated binding \mcoreinline{a} on lines~\ref{lst:lamlift2:acc1}-\ref{lst:lamlift2:acc2} in Listing~\ref{lst:lamlift2}. We refer to the expression assigned to an identifier of a let-expression as the body. For instance, the body of \mcoreinline{a} is the expression starting after the equals sign on line~\ref{lst:lamlift2:acc1} until the \mcoreinline{in} keyword at the end of line~\ref{lst:lamlift2:acc2}.

Because we perform lambda lifting to capture free variables of non-function type, we know any free variables in the body of \mcoreinline{a} must be functions. These functions are what \mcoreinline{a} depends on and, therefore, what we must include in the extracted program. For instance, we find that \mcoreinline{f} (on line~\ref{lst:lamlift2:f-free}) is a free variable, while \mcoreinline{map} and \mcoreinline{reduce} are parallel keywords and \mcoreinline{addi} is an intrinsic for integer addition. We include the body of \mcoreinline{f} in the extracted code, and we consider its body in the same manner to find transitive dependencies. This is repeated for the remainder of the program, eventually resulting in the highlighted parts of Listing~\ref{lst:lamlift2} being extracted.

\subsection{Classification}
\label{subsec:classification}

The extraction produces the accelerated code. We use the classification to determine, for each accelerated expression, which backend to compile it with. As discussed in Section~\ref{subsec:key-idea}, we define functional parallel expressions (for the Futhark backend) and imperative parallel extensions (for the CUDA backend). The classifier determines which backend to use for each accelerated expression based on the parallel extensions we use within it. This approach enables using both backends within one program (as in Listing~\ref{lst:mot-ex}). We reuse the extraction approach to produce an expression for each target backend to ensure that these expressions only include what is required.

\subsection{Marshaling}
\label{subsec:marshaling}

The data representation used in the sequential code is distinct from the accelerated code. In the backend compiler phase, data marshaling code is automatically generated based on the types of accelerated bindings. In the sequential code, we replace the accelerate bindings (e.g., \mcoreinline{a} on lines~\ref{lst:lamlift2:acc1}-\ref{lst:lamlift2:acc2} of Listing~\ref{lst:lamlift2}) with an external function running the backend-specific code. The data marshaling is performed by converting the arguments of an accelerated binding to the backend-specific format (\mcoreinline{s} and \mcoreinline{c} on line~\ref{lst:lamlift2:sum} in Listing~\ref{lst:lamlift2}) and, similarly, to convert the resulting value from the backend to the format used in the sequential code (the value assigned to \mcoreinline{sum} on line~\ref{lst:lamlift2:sum}).

\subsection{Well-formedness}
\label{subsec:well-formedness}

When generating accelerated code, the target backend is typically more restrictive than the default sequential compiler (which, for instance, uses a garbage collector). We perform well-formedness checks to ensure the accelerated code adheres to our defined rules for the CUDA and Futhark backends. The well-formedness includes both assumptions we check at runtime (in debug mode) and static rules we verify at compile-time (the well-formedness checker). For instance, our well-formedness rules ensure that the accelerated code does not pass higher-order functions as arguments, as this is not supported by either accelerate backend.

\section{Evaluation}
\label{sec:evaluation}

In this section, we evaluate the compiler. For the Futhark backend, we implement a small Futhark benchmark suite in the parallel language (Section~\ref{subsec:futhark-evaluation}). For the CUDA backend, we implement an ODE solver and a neural network (Section~\ref{subsec:cuda-evaluation}). These benchmarks show that our compiler for expression acceleration can solve non-trivial problems and significantly improve performance over sequential code.

We run the experiments on an Intel Xeon 656 Gold 6136 CPU and a Titan RTX GPU using Ubuntu 18.04. We use OCaml version 4.14.0, CUDA version 11.7, and Futhark version 0.25.16. The compilation times for all benchmarks presented in this section are below $5$ seconds on this machine.

\subsection{Futhark}
\label{subsec:futhark-evaluation}

To show that our compiler can generate efficient code, we implement the Parboil suite of the Futhark benchmark suite\footnote{https://github.com/diku-dk/futhark-benchmarks} in our parallel language and compare the Futhark output from our compiler with existing code using the built-in benchmarking tool in the Futhark compiler (\texttt{futhark bench}) using Futhark's CUDA backend, measuring execution time and validating the output for each dataset.

To be able to make a comparison using \texttt{futhark bench}, we post-process the Futhark output from our compiler. We add metadata used by the benchmarking tool, and for compatibility in a few of the benchmarks, we add a custom entry function to convert integer types as our compiler always uses 64-bit integers. Importantly, these post-processing steps are not required to use the binary produced by our compiler. For this evaluation, we use a keyword for specifying inline Futhark expressions in the accelerated code.

\begin{table}[t!]
\centering
\begin{tabular}{lll}
Dataset & Accelerate & Futhark\\
\hline
\texttt{histo-default} & $271.5 \pm 3.4$ & $240.7 \pm 2.9$\\
\texttt{histo-large} & $215.6 \pm 3.3$ & $202.0 \pm 3.4$\\
\texttt{mri-q-large} & $2695.3 \pm 36.4$ & $2739.9 \pm 39.4$\\
\texttt{mri-q-small} & $581.7 \pm 4.2$ & $591.4 \pm 4.3$\\
\texttt{sgemm-medium} & $964.6 \pm 1.9$ & $978.7 \pm 2.2$\\
\texttt{sgemm-small} & $48.6 \pm 1.7$ & $49.7 \pm 1.6$\\
\texttt{sgemm-tiny} & $22.8 \pm 4.3$ & $25.4 \pm 3.2$\\
\texttt{stencil-default} & $31838.0 \pm 317.0$ & $36995.0 \pm 154.6$\\
\texttt{stencil-small} & $987.6 \pm 10.3$ & $1041.8 \pm 9.7$\\
\texttt{tpacf-large} & $758541.1 \pm 1833.8$ & $745027.7 \pm 2219.1$\\
\texttt{tpacf-medium} & $126135.6 \pm 588.4$ & $124062.4 \pm 6.2$\\
\texttt{tpacf-small} & $1795.3 \pm 2.8$ & $1802.0 \pm 31.3$\\
\end{tabular}
\caption{Execution times (in microseconds) as reported by \texttt{futhark bench}. Each dataset is identified by the benchmark name and the name of the input data used for that run.}
\label{table:futhark-evaluation}
\end{table}

Our results exclude the \texttt{lbm} benchmark because the reference implementation relies on \emph{ownership types} to achieve high performance. This is a type constraint in Futhark not present in our source language, and this concept is required to achieve high performance. We present the outcome for the other benchmarks in Table~\ref{table:futhark-evaluation}. The performance of the accelerated code (\texttt{Accelerate}) is on par with the existing implementations (\texttt{Futhark}) for most benchmarks.

The differences in performance are mainly because our compiler always uses 64-bit integers. This is for simplicity; the source language only includes one integer type, and integers have to be 64-bit in certain situations in Futhark. For most benchmarks where this matters, it puts our code at a slight disadvantage (e.g., the \texttt{histo} benchmarks). However, the Futhark code produced by our compiler is noticeably faster for the \texttt{stencil} benchmarks. We found two reasons for this. First, the reference implementation frequently converts between 32- and 64-bit integers, leading to a slight loss of performance according to our testing. Second, our version does not short-circuit conditional expressions, as in the reference implementation. This seems to have a rather big impact on the performance, according to our measurements.

\subsection{CUDA}
\label{subsec:cuda-evaluation}

\begin{table}[t!]
\centering
\begin{tabular}{l|ll}
Version & Sequential & Accelerated\\
\hline
ODE-$10^3$ & $7.67 \pm 0.05$ & $0.30 \pm 0.01$\\
ODE-$10^4$ & $76.66 \pm 0.37$ & $0.54 \pm 0.02$\\
ODE-$10^5$ & $768.45 \pm 6.62$ & $0.74 \pm 0.01$\\
NN-MNIST & $649.81 \pm 7.63$ & $0.50 \pm 0.01$\\
NN-CIFAR-10 & $12981.30 \pm 120.42$ & $3.38 \pm 0.03$
\end{tabular}
\caption{Execution times (in seconds) for benchmarks using CUDA acceleration. We run three ODE benchmarks and two neural network benchmarks, ordered by increasing problem size.}
\label{table:cuda-evaluation}
\end{table}

We present execution-time results for the ODE solver and the neural networks in Table~\ref{table:cuda-evaluation}. For these benchmarks, we accelerate using the CUDA backend and compare the performance of the code when compiled without and with acceleration enabled (labeled Sequential and Accelerated, respectively). We measure the wall time of executing the accelerated expression. This includes overheads due to marshaling and copying between the CPU and the GPU but excludes I/O.

We make two key observations when comparing the sequential and the accelerated results. First, unsurprisingly, the accelerated versions have better performance overall. We also observe that they scale better when the problem size increases; this is because the smaller benchmarks are not fully utilizing the GPU. Second, we see that the overhead of marshaling and copying data between the CPU and the GPU is negligible compared to the performance gains, in these cases. Expression acceleration works best when most work is focused on a single or few uses of \mcoreinline{accelerate}, as this reduces the overheads of marshaling and copying.

\section{Related Work}
\label{sec:related-work}

Our approach is similar to dynamic approaches in Python~\cite{catanzaro2011copperhead,zhou2024appy,team2016theano,lam2015numba,tillet2019triton,frostig2018compiling}, in particular Numba~\cite{lam2015numba} and JAX~\cite{frostig2018compiling}, where functions are annotated to indicate acceleration. Expression acceleration is a static approach, which requires static analyses to extract code to different backends and ensure well-formedness.

Similar approaches include standalone languages~\cite{henriksen2017futhark, leissa2018anydsl, dubach2012compiling}, domain-specific languages embedded in general-purpose programming languages~\cite{steuwer2017lift, hagedorn2020achieving, sujeeth2014delite, svensson2008obsidian, chakravarty2011accelerating, kelley2013halide}, library APIs~\cite{ishizaki2015compiling, rossbach2013dandelion, EnmurenEtAl:2010, ErnstssonEtal:2018}, and annotation-based approaches in low-level languages~\cite{openacc, openmp}. Our work is distinguished by the simplicity of \mcoreinline{accelerate}, which allows us to seamlessly move parts of a program to the GPU without having to modify other parts. For instance, extraction ensures all dependencies are part of the GPU code, and the compiler automatically handles the data marshaling. Furthermore, as this approach is implemented in the compiler, many classes of errors can be detected at compile-time rather than at runtime.

\section{Conclusions}
\label{sec:conclusions}

This paper presents our compiler implementation of \emph{expression acceleration} for a high-level typed programming language. The key ideas include expression extraction, classification, well-formedness checking, and compilation targeting multiple backends. The fundamental challenge of this approach is making it expressive and supporting a large subset of the host language. The evaluation shows that our compiler supports non-trivial examples and that it can significantly improve performance over fully sequential code.

\begin{acks}

We want to thank Linnea Stjerna, Viktor Palmkvist, and Anders Ågren Thuné for their assistance during development.

This project is financially supported by the Swedish Foundation for Strategic Research (FFL15-0032 and RIT15-0012). The research has also been carried out as part of the Vinnova Competence Center for Trustworthy Edge Computing Systems and Applications (TECoSA) at the KTH Royal Institute of Technology.

\end{acks}

\bibliography{references,dbro,phaller}

\appendix

\input{appendix}

\end{document}

%% file: mcorelisting.tex
\usepackage{listings}
\usepackage{xcolor}

\def \mcorefontsize {\footnotesize}

\def \mcorefont {\fontfamily{pcr}\selectfont\mcorefontsize}

\definecolor{ckeywords}{rgb}{0.13,0.13,1}
\definecolor{ccomments}{rgb}{0,0.5,0.5}
\definecolor{cstrings}{rgb}{0,0.5,0}
\definecolor{cwarnings}{rgb}{1,0.5,0}

\lstdefinelanguage{MCore}{
    morekeywords={Lam,con,else,end,fix,if,in,lam,lang,let,match,recursive,sem,syn,then,type,use,utest,with,accelerate,map,reduce,loop,map2,flatten},
    otherkeywords={->},
    keywordstyle=\color{ckeywords},
    morekeywords=[2]{mexpr,include,never},
    keywordstyle=[2]\color{cwarnings},
    morecomment=[l][\color{ccomments}]{--},
    morestring=[b]",
    stringstyle=\color{cstrings},
    sensitive=true,
    basicstyle=\mcorefont,
    breaklines=true,
    escapeinside={(*@}{@*)},
    numbers=left,
    stepnumber=1,
    numberstyle=\tiny,
    xleftmargin=1em,
    columns=fixed, 
    showstringspaces=false,
    mathescape=true,
    breaklines=true,
    breakatwhitespace=true,
    mathescape=true,
    showstringspaces=false
  }

\lstnewenvironment{mcore}
  {
    \lstset{
      language=MCore,
      numbers=none,
      xleftmargin=0pt
    }
  }{}

\lstnewenvironment{mcore-lines}
  {
    \lstset{
      language=MCore,
      xleftmargin=0pt
    }
  }{}

\newcommand{\mcoreinline}[1]{\lstinline[{language=MCore}]|#1|}

%% file: commands.tex
\newcommand{\msf}[1]{\mathtt{#1}}
\newcommand{\pluseq}{\mathrel{+}=}

\newcommand{\accelerate}[1]{\msf{accelerate}\ #1}

\newcommand{\letexpr}[3]{\msf{let}\ #1 = #2\ \msf{in}\ #3}
\newcommand{\reclet}[3]{\msf{recursive}\ {(\msf{let}\ #1_i = #2_i)}^{i \in 1 \ldots n}\ \msf{in}\ #3}

\newcommand{\matchexpr}[4]{
\msf{match}\ #1\ \msf{with}\ #2\ \msf{then}\ #3\ \msf{else}\ #4}
\newcommand{\loopexpr}[2]{\msf{loop}\ #1\ #2}
\newcommand{\mapexpr}[2]{\msf{map}\ #1\ #2}
\newcommand{\maptexpr}[3]{\msf{map2}\ #1\ #2\ #3}
\newcommand{\reduceexpr}[3]{\msf{reduce}\ #1\ #2\ #3}
\newcommand{\flattenexpr}[1]{\msf{flatten}\ #1}

\newcommand{\fv}[1]{\msf{FV}(#1)}
\newcommand{\boundp}[1]{\msf{B}(#1)}

\newcommand{\typedlamexpr}[3]{\lambda #1 : #2 .\ #3}
\newcommand{\typedletexpr}[4]{\msf{let}\ #1 : #2 = #3\ \msf{in}\ #4}
\newcommand{\typedreclet}[4]{\msf{recursive}\ {(\msf{let}\ #1_i : #2_i = #3_i)}^{i \in 1 \ldots n}\ \msf{in}\ #4}

%% file: appendix.tex
\section{Source Language}
\label{sec:appendix:source-language}

We use the Miking framework's intermediate language called MExpr as a basis of our source language. MExpr is a simple yet complete typed functional core language that is easily extensible. We use a subset of the core language (Figure~\ref{fig:mexpr-ast}). The language consists of standard functional language expressions, such as lambdas, let-expressions, applications, and recursive bindings.

\begin{figure}[t!]
\[
\begin{array}{ll}
x & \text{variables}\\
n & \text{natural numbers}\\
l & \text{record labels}\\
c & \text{constant values}\\
T_g & \text{ground types}\\
\end{array}
\]
\[
\begin{array}{lcl}
e & ::= & \lambda x : T. e\\
  &   | & \msf{let}\ x : T = e_1\ \msf{in}\ e_2\\
  &   | & \msf{recursive}\ {(\msf{let}\ x_i : T_i = e_i)}^{i \in 1 \ldots n}\ \msf{in}\ e\\
  &   | & \msf{match}\ e_1\ \msf{with}\ p\ \msf{then}\ e_2\ \msf{else}\ e_3\ |\ \msf{never}\\
  &   | & e_1\ e_2\ |\ \{l_i = e_i\}^{i \in 1 \ldots n}\ |\ [e_i]^{i \in 1 \ldots n}\ |\ x\ |\ c\\
p & ::= & \{l_i = p_i\}^{i \in 1 \ldots n}\ |\ x\ |\ c\\
T & ::= & T_g\ |\ T_1 \rightarrow T_2\ |\ \{l_i : T_i\}^{i \in 1 \ldots n}\ |\ [T]\ |\ \msf{Tensor}[T]\\
\end{array}
\]
\caption{Definition of the base language, a subset of the MExpr language.}
\label{fig:mexpr-ast}
\end{figure}

Records are represented as $\{l_i = e_i\}^{i \in 1 \ldots n}$ where $l_i$ is a label, and $e_i$ is an expression bound to that label. Sequence literals are represented as $[e_i]^{i \in 1 \ldots n}$, where each expression $e_i$ represents an element in the sequence. Patterns $p$ consist of variables $x$ and constants $c$, as well as record patterns $\{l_i = p_i\}^{i \in 1 \ldots n}$. In the language of Figure~\ref{fig:mexpr-ast}, sequences are immutable containers, while tensors are mutable containers with a dynamic number of dimensions. Sequences are constructed using a literal value, but there is no literal for tensors. Tensors are instead constructed from built-in functions. Types are defined closely following the same style as expressions and patterns. For example, a sequence containing elements of type $T$ has type $[T]$ and a tensor with elements of type $T$ has
type $\msf{Tensor}[T]$.

Literal values of ground types $T_g$, such as integers and floating-point numbers, are included in the constants $c$. Constants also include built-in curried functions such as
\[
c \in \{\msf{addi}, \msf{muli}, \msf{create}, \msf{tensorCreate}, \ldots\}
\]
The MExpr language of Figure~\ref{fig:mexpr-ast} does not include the parallel operations we need for acceleration. We present a subset of the parallel extensions we make in Figure~\ref{fig:pmexpr-extension}. In~\eqref{eq:pmexpr-ast}, we define the \mcoreinline{accelerate} keyword for controlling which expressions are accelerated. We also need expressions that evaluate in parallel when used in accelerated code. Therefore, we also define extensions for the Futhark and CUDA backends.

\begin{figure}[t!]
\begin{alignat}{3}
&\begin{aligned}
e \quad && \pluseq \quad & \accelerate{e}
\end{aligned}\label{eq:pmexpr-ast}\\
&\begin{aligned}
e \quad && \pluseq \quad & \mapexpr{e_1}{e_2}\\[-3pt]
  \quad && |       \quad & \maptexpr{e_1}{e_2}{e_3}\\[-3pt]
  \quad && |       \quad & \reduceexpr{e_1}{e_2}{e_3}\\[-3pt]
  \quad && |       \quad & \flattenexpr{e_1}\\[-3pt]
\end{aligned}\label{eq:futhark-ast}\\
&\begin{aligned}
e \quad && \pluseq \quad & \loopexpr{e_1}{e_2}
\end{aligned}\label{eq:cuda-ast}
\end{alignat}
\caption{Extensions of expressions $e$ in the sequential core language. Our source language is the union of these extensions with the source language.}
\label{fig:pmexpr-extension}
\end{figure}

We define the key parallel expressions used in the Futhark backend in~\eqref{eq:futhark-ast} of Figure~\ref{fig:pmexpr-extension}. The \mcoreinline{map} expression takes a function expression $e_1$ and applies it to the elements of a sequence $e_2$. Similarly, \mcoreinline{map2} applies a function $e_1$ to the elements of two sequences $e_2$ and $e_3$. For example, we compute the elementwise sum of two sequences in parallel as
\begin{center}
\mcoreinline{accelerate (map2 addi s1 s2)}
\end{center}

The \mcoreinline{reduce} expression takes a function expression $e_1$, the initial value of the accumulator $e_2$, and a sequence $e_3$ to operate on. For example, we can use it to compute the product of a sequence \mcoreinline{s} of integers in parallel as
\begin{center}
\mcoreinline{accelerate (reduce muli 1 s)}
\end{center}

The \mcoreinline{flatten} expression translates a two-dimensional sequence to a one-dimensional sequence by concatenating the inner sequences. In addition to the parallel extensions for the Futhark backend, we also include a keyword for writing inline Futhark code. This is used in the evaluation (\Cref{subsec:futhark-evaluation}) to speed up development.

In~\eqref{eq:cuda-ast} of Figure~\ref{fig:pmexpr-extension}, we show an extension used for the CUDA backend. The \mcoreinline{loop} expression corresponds to a for-loop, executed in parallel when used within accelerated code. It takes an integer argument $e_1$ denoting the iteration count and an iteration function $e_2$ corresponding to the loop body. The function $e_2$ is invoked once for each integer in the range $[0, e_1)$, in an undefined order. As tensors are mutable, we can encode operations over multiple tensors using a loop. Given tensors \mcoreinline{a} and \mcoreinline{b} of length \mcoreinline{n}, we use a loop to compute the elementwise product $\msf{b} = 2 \cdot \msf{a}$ in parallel as
\begin{mcore}
let f = lam i.
  tensorSet b [i] (muli 2 (tensorGet a i)) in
accelerate (loop n f)
\end{mcore}
The constant \mcoreinline{muli} performs integer multiplication, while \mcoreinline{tensorGet} and \mcoreinline{tensorSet} read from and write to tensors.

We construct the parallel source language PMExpr by combining the source language of Figure~\ref{fig:mexpr-ast} with the extensions of Figure~\ref{fig:pmexpr-extension}. This enables each accelerated expression to use either backend, meaning both backends may be used in the same program. For example, the program of Listing~\ref{lst:mot-ex} uses the CUDA backend on lines~\ref{lst:mot-ex:cuda-acc1}-\ref{lst:mot-ex:cuda-acc2}, and the Futhark backend on line~\ref{lst:mot-ex:fut-acc}.

\begin{figure*}[t!]
\small
\begin{subfigure}{.30\textwidth}
\begin{align*}
\boundp{x} &= \{x\}\\
\boundp{c} &= \emptyset\\
\boundp{\{l_i = p_i\}^{i \in 1 \ldots n}} &= \bigcup_{i \in 1 \ldots n} \boundp{p_i}
\end{align*}
\caption{Definition of the $\boundp{p}$ function, computing the set of variables bound by patterns.}
\label{fig:boundp}
\end{subfigure}
\hfill
\begin{subfigure}{.64\textwidth}
\centering
\begin{align*}
&\fv{x} = \{x\}\\
&\fv{c} = \emptyset\\
&\fv{\typedlamexpr{x}{T}{e}} = \fv{e} \setminus \{x\}\\
&\fv{e_1\ e_2} = \fv{e_1} \cup \fv{e_2}\\
&\fv{\typedletexpr{x}{T}{e_1}{e_2}} = \fv{e_1} \cup (\fv{e_2} \setminus \{x\})\\
&\fv{\typedreclet{x}{T}{e}{e}} = \left(\bigcup_{i \in 1 \ldots n} \fv{e_i} \cup \fv{e}\right) \setminus \{x_i\ |\ i \in 1 \ldots n\}\\
&\fv{\matchexpr{e_1}{p}{e_2}{e_3}} = \fv{e_1} \cup (\fv{e_2} \setminus \boundp{p}) \cup \fv{e_3}\\
&\fv{\msf{never}} = \emptyset\\
&\fv{\{l_i = e_i\}^{i \in 1 \ldots n}} = \bigcup_{i \in 1 \ldots n} \fv{e_i}\\
&\fv{[e_i]^{i \in 1 \ldots n}} = \bigcup_{i \in 1 \ldots n} \fv{e_i}\\
&\fv{\accelerate{e}} = \fv{e}\\
&\fv{\mapexpr{e_1}{e_2}} = \fv{e_1} \cup \fv{e_2}\\
&\fv{\maptexpr{e_1}{e_2}{e_3}} = \fv{e_1} \cup \fv{e_2} \cup \fv{e_3}\\
&\fv{\reduceexpr{e_1}{e_2}{e_3}} = \fv{e_1} \cup \fv{e_2} \cup \fv{e_3}\\
&\fv{\flattenexpr{e_1}} = \fv{e_1}\\
&\fv{\loopexpr{e_1}{e_2}} = \fv{e_1} \cup \fv{e_2}
\end{align*}
\caption{Definition of the $\fv{e}$ function for computing the free variables of an expression $e$.}
\label{fig:freevars}
\end{subfigure}
\caption{Definition of functions \texttt{B} and \texttt{FV}.}
\label{fig:freevar-top}
\end{figure*}

\section{Free variables}
\label{sec:appendix:free-variables}

We present the complete definition of free variables for all PMExpr expressions in Figure~\ref{fig:freevar-top}. In Figure~\ref{fig:boundp}, we define the supporting function $\boundp{p}$, which computes the set of variables bound in a pattern $p$. Based on this definition, we present the complete definition of $\fv{e}$ computing the set of free variables in an expression $e$ in Figure~\ref{fig:freevars}. This definition is used in the definitions of classification in Section~\ref{sec:appendix:classification} and in the well-formedness rules in Section~\ref{sec:appendix:well-formedness}.

\section{Extraction}
\label{sec:appendix:extraction}

In this section, we give a formal presentation of the extraction in terms of the \texttt{Extract} function of Algorithm~\ref{alg:extract}. In the algorithm, we use \texttt{typewriter font} to refer to AST nodes. The function takes a set of identifiers $I$, which correspond to the identifiers of accelerated bindings that are to be extracted. It also takes an expression $e$ to extract from. We use the term \emph{binding} to refer to a let-expression or a binding in a recursive let-expression. The result consists of two parts. The first is an updated set $I'$, containing the identifiers of all extracted bindings. The second part is an expression $e'$, which corresponds to the extracted program. For instance, when Algorithm~\ref{alg:extract} is applied to Listing~\ref{lst:lamlift2} (giving identifiers $I_{\msf{acc}} = \{\texttt{a}\}$ and the whole program expression as input), the returned identifiers are $I' = \{\texttt{a}, \texttt{f}, \texttt{g}\}$, and the returned expression $e'$ is equal to the highlighted parts of Listing~\ref{lst:lamlift2}.

\begin{algorithm}[t!]
\caption{Extraction of accelerated code given an input program and a set of
identifiers corresponding to \mcoreinline{accelerate} expressions.}
\label{alg:extract}
\begin{algorithmic}[1]
\State $I$: identifiers to be included in the extracted expression
\State $e$: the lambda lifted source expression to extract from
\Function{Extract}{$I, e$}
  \Match{$e$}{$\reclet{x}{e}{e_0}$} \label{alg:extract:reclet1}
    \State $(I', e'_0) \gets \Call{Extract}{I, e_0}$ \label{alg:extract:reccall1}
    \State $B \gets \{x_i\ |\ i \in 1 \ldots n\}$ \label{alg:extract:recletb}
    \State $I_B \gets \{x_i\ |\ x_i \in B \cap I'\}$ \label{alg:extract:recletib}
    \State $I_B \gets \msf{FIX}_{I_B} \left(I_B \cup \left(\left(\bigcup_{x_i \in I_B} \fv{e_i}\right) \cap B\right)\right)$ \label{alg:extract:recletids}
    \State $I'' \gets I' \cup I_B \cup \left(\bigcup_{x_i \in I_B} \fv{e_i}\right)$ \label{alg:extract:iupd}
    \State $N \gets \{i\ |\ x_i \in I_B\}$ \label{alg:extract:ndef}
    \State $e' \gets \msf{recursive}\ {(\msf{let}\ x_j = e_j)}^{j \in N}\ \msf{in}\ e'_0$ \label{alg:extract:recletext}
    \State \textbf{return} $(I'', e')$ \label{alg:extract:reclet2}
  \MatchEls{$e$}{$\letexpr{x}{e_1}{e_2}$} \label{alg:extract:let1}
    \State $(I', e_2') \gets \Call{Extract}{I, e_2}$ \label{alg:extract:reccall2}
    \If{$x \in I'$} \label{alg:extract:ins}
      \State \textbf{return} $(I' \cup \fv{e_1}, \letexpr{x}{e_1}{e_2'})$ \label{alg:extract:thnins}
    \Else\ \textbf{return} $(I', e_2')$ \label{alg:extract:notins}
    \EndIf \label{alg:extract:let2}
  \MatchEnd{\textbf{return} $(I, \{\})$} \label{alg:extract:basecase}
\EndFunction
\end{algorithmic}
\end{algorithm}

The \texttt{Extract} function in Algorithm~\ref{alg:extract} consists of three cases, by matching on the shape of the input expression \mcoreinline{e}. The first case handles recursive let-expressions (line~\ref{alg:extract:reclet1}) and the second case handles let-expressions (line~\ref{alg:extract:let1}). Note that both these cases start with a recursive call to the \texttt{Extract} function on the in-expression of the recursive let- and let-expressions, respectively. That is, the algorithm traverses the program in a bottom-up fashion. The final case, which is also the base case of the function, matches other kinds of expressions (line~\ref{alg:extract:basecase}).

Consider the program of Listing~\ref{lst:lamlift2} presented in Section~\ref{subsec:extraction}. In this program, we only have one accelerated binding \mcoreinline{a} to extract, on lines~\ref{lst:lamlift2:acc1}-\ref{lst:lamlift2:acc2}. Therefore the input set $I_{\msf{acc}}$ is $\{a\}$. For a program with multiple \mcoreinline{accelerate} expressions, the input set $I_{\msf{acc}}$ consists of multiple identifiers. The complete program of Listing~\ref{lst:lamlift2} corresponds to the argument $e$ to \texttt{Extract}.

As we noted, the \texttt{Extract} function traverses the program in a bottom-up fashion, as the let- and recursive let-expression cases start with a self-recursive call. Thus, we reach the base case with the \mcoreinline{printSum sum} expression on line~\ref{lst:lamlift2:print}. For this expression, we return $I$ as is, and an empty record $\{\}$ as the extracted expression. The next expression to consider is that of line~\ref{lst:lamlift2:sum} in Listing~\ref{lst:lamlift2}. This is a let-expression, so we are in the second case of \texttt{Extract} (line~\ref{alg:extract:let1}). In the recursive call on line~\ref{alg:extract:reccall2}, we get the returned values from the base case. On line~\ref{alg:extract:ins}, we check if the identifier $\msf{sum}$ is in $I'$. As $I$ was returned in the base case, the value of $I'$ is $\{\msf{a}\}$. Therefore the condition evaluates to false. Thus, we return the $I'$ without modifying it, and $\{\}$ from the base case.

Next, we consider the let-expression on lines~\ref{lst:lamlift2:acc1}-\ref{lst:lamlift2:acc2} in Listing~\ref{lst:lamlift2}. We enter the let-expression case of the \texttt{Extract} function once more. The recursive call results in $I'$ equal to $I$ and $e_2'$ equal to $\{\}$, as was the result of the previous case. This time, the condition on line~\ref{alg:extract:ins} of \texttt{Extract} evaluates to true, as $I' = I = \{\msf{a}\}$. The result on line~\ref{alg:extract:thnins} is computed as follows. We update the set of identifiers $I$ by including the free variables in the body of the let-expression, $e_1$. We find that $\fv{e_1} = \{\msf{f}\}$ in this case, and thus we return $I = \{\msf{a}, \msf{f}\}$.

The end result of applying \texttt{Extract} on Listing~\ref{lst:lamlift2} is the set of identifiers $\{\msf{a}, \msf{f}, \msf{g}\}$. The extracted expression includes the highlighted expressions of Listing~\ref{lst:lamlift2}, with a trailing $\{\}$ as the final in-expression at the end of line~\ref{lst:lamlift2:acc2}.

\section{Classification}
\label{sec:appendix:classification}

The extraction produces the accelerated code, containing the accelerated bindings and the bindings they depend on. By classifying the accelerated code, the compiler decides which backend to use for each accelerated binding.

The parallel expressions we introduce may be used in either one of the backends. Therefore, we classify expressions based on which parallel expressions are used. This enables efficient production of backend-specific programs from accelerated code. For example, consider the program of Listing~\ref{lst:lamlift2}. The accelerated binding on lines~\ref{lst:lamlift2:acc1}-\ref{lst:lamlift2:acc2} uses the \mcoreinline{map} and \mcoreinline{reduce} expressions. These are extensions for the Futhark backend, so we use that backend to compile the binding.

\begin{figure}[t!]
\begin{subfigure}{.5\textwidth}
\[
\begin{aligned}
&P_F(e) & = |\{e'\ |&\ e' \in S(e) \land\\
&                 &           & (e' \equiv \mapexpr{e_1}{e_2} \lor e' \equiv \maptexpr{e_1}{e_2}{e_3} \lor\\
&                 &           &  e' \equiv \reduceexpr{e_1}{e_2}{e_3} \lor e' \equiv \flattenexpr{e_1})\}|\\
&P_C(e)  & = |\{e'\ |&\ e' \in S(e) \land e' \equiv \loopexpr{e_1}{e_2}\}|
\end{aligned}
\]
\caption{Definition of $P_F$ and $P_C$, computing the
number of occurrences of parallel expressions in a given expression $e$.}
\label{fig:sexprdef}
\end{subfigure}
\begin{subfigure}{.5\textwidth}
\begin{align*}
P^V_F(x) &=
\begin{cases}
P_F(e_x) + \sum\limits_{y \in \fv{e_x}} P^{V \cup \{x\}}_F(y) & \text{if }x \not\in V\\
0 & \text{if }x \in V
\end{cases}\\
P^V_C(x) &=
\begin{cases}
P_C(e_x) + \sum\limits_{y \in \fv{e_x}} P^{V \cup \{x\}}_C(y) & \text{if }x \not\in V\\
0 & \text{if }x \in V
\end{cases}
\end{align*}
\caption{Definition of the functions $P^V_F$ and $P^V_C$, which compute the number of occurrences of parallel expressions in a binding $x$ with body $e_x$, given a set of visited identifiers $V$.}
\label{fig:sbinddef}
\end{subfigure}
\begin{subfigure}{.5\textwidth}
\begin{displaymath}
C(x) =
\begin{cases}
  \msf{Any} & \text{if }P^{\emptyset}_F(x) = 0 \land P^{\emptyset}_C(x) = 0\\
  \msf{Futhark} & \text{if }P^{\emptyset}_F(x) > 0 \land P^{\emptyset}_C(x) = 0\\
  \msf{CUDA} & \text{if }P^{\emptyset}_F(x) = 0 \land P^{\emptyset}_C(x) > 0\\
  \msf{Invalid} & \text{if }P^{\emptyset}_F(x) > 0 \land P^{\emptyset}_C(x) > 0
\end{cases}
\end{displaymath}
\caption{Definition of the binding classification function $C$.}
\label{fig:classify}
\end{subfigure}
\caption{Helper definitions used to define the classification.}
\label{fig:class-defns}
\end{figure}

The input to the classification is the accelerated code $e_{\msf{acc}}$ produced by the extraction. The output is two programs, one for the Futhark backend and another for the CUDA backend. Figure~\ref{fig:class-defns} contains the formal definitions we use for the classification. Also, we define a function $S$ for computing the set of all subexpressions of an expression $e$, including $e$. For brevity, we define $S$ in terms of an auxiliary function $S'$ which excludes $e$ (i.e., $S(e) = S'(e) \cup \{e\}$) in Figure~\ref{fig:subexpr}. Note that this approach to classification is for presentation purposes only --- in the implementation, we perform classification by traversing the AST more efficiently.

\begin{figure}[t!]
\footnotesize
\[
\begin{array}{l}
S'(\accelerate{e}) = S(e)\\
S'(\typedlamexpr{x}{T}{e}) = S(e)\\
S'(\typedletexpr{x}{T}{e_1}{e_2}) = S(e_1) \cup S(e_2)\\
S'(\typedreclet{x}{T}{e}{e}) = (\bigcup_{i \in 1 \ldots n} S(e_i)) \cup S(e)\\
S'(\matchexpr{e_1}{p}{e_2}{e_3}) = S(e_1) \cup S(e_2) \cup S(e_3)\\
S'(\msf{never}) = \emptyset\\
S'(e_1\ e_2) = S(e_1) \cup S(e_2)\\
S'(\{l_i : e_i\}^{i \in 1 \ldots n}) = \bigcup_{i \in 1 \ldots n} S(e_i)\\
S'([e_i]^{i \in 1 \ldots n}) = \bigcup_{i \in 1 \ldots n} S(e_i)\\
S'(x) = \emptyset\\
S'(c) = \emptyset\\
S'(\msf{map}\ e_1\ e_2) = S(e_1) \cup S(e_2)\\
S'(\msf{map2}\ e_1\ e_2\ e_3) = S(e_1) \cup S(e_2) \cup S(e_3)\\
S'(\msf{reduce}\ e_1\ e_2\ e_3) = S(e_1) \cup S(e_2) \cup S(e_3)\\
S'(\msf{flatten}\ e_1) = S(e_1)\\
S'(\msf{inlineFuthark}\ e_1) = \emptyset\\
S'(\msf{loop}\ e_1\ e_2) = S(e_1) \cup S(e_2)\\
\end{array}
\]
\caption{Definition of the self-excluding subexpression function $S'$ applied to an expression $e$.}
\label{fig:subexpr}
\end{figure}

In Figure~\ref{fig:sexprdef}, we define the $P_F$ and $P_C$ functions. These take an expression $e$ as input and compute the number of occurrences of parallel expressions for Futhark or CUDA, respectively. We use $e' \equiv p$ to denote pattern matching. This evaluates to true if $e'$ matches the pattern expression $p$. For instance, if $e' = \msf{map}\ f\ s$, then $e' \equiv \msf{map}\ e_1\ e_2$ is true.

We define similar functions $P_F^V$ and $P_C^V$ on bindings in Figure~\ref{fig:sbinddef}. A binding has an identifier $x$ and it is bound to an expression $e_x$ (e.g., $\msf{let}\ x = e_x\ \msf{in} \ldots$). For a given identifier of a binding $x$, we assume that we know its bound expression $e_x$. We compute the number of occurrences of parallel expressions in a binding in two steps. First, we consider the occurrences in its body $e_x$, using $P_F$ or $P_C$. Second, we include occurrences in expressions of bindings that $e_x$ depends on, i.e., given by the free variables in $e_x$. We keep track of the visited bindings in the set $V$ of identifiers to prevent bindings from being counted more than once.

We define the binding classification function $C$ in Figure~\ref{fig:classify}. This function takes a binding as input and categorizes it as an element of the set $\{\msf{Any}, \msf{Futhark}, \msf{CUDA}, \msf{Invalid}\}$. The function $C$ is used to construct two subsets of identifiers, one for each backend, as
\begin{align*}
I_F &= \{a\ |\ a \in I_{\msf{acc}} \land C(a) = \msf{Futhark}\}\\
I_C &= \{a\ |\ a \in I_{\msf{acc}} \land C(a) = \msf{CUDA}\}
\end{align*}

Recall that $I_{\msf{acc}}$ is the set of identifiers of all accelerated bindings in the program. The compilation fails if an accelerated binding $a$ is classified as \texttt{Any} or \texttt{Invalid}. In the first case, $a$ does not use any parallel expressions, which goes against the purpose of using acceleration. In the second case, $a$ uses parallel expressions of both backends, which is unsupported.

We have now classified the identifiers used for the Futhark and CUDA backend, $I_F$ and $I_C$, respectively. However, at the end of this phase, we need the backend-specific programs $e_F$ and $e_C$. To produce these two backend-specific programs, we reuse the \texttt{Extract} function of Algorithm~\ref{alg:extract} as follows
\begin{align*}
(I_F', e_F) &\leftarrow \msf{Extract}(I_F, e_{\msf{acc}})\\
(I_C', e_C) &\leftarrow \msf{Extract}(I_C, e_{\msf{acc}})
\end{align*}

We use $e_{\msf{acc}}$ to denote the accelerated code produced by the accelerate
extraction in Section~\ref{subsec:extraction}.

\section{Well-formedness}
\label{sec:appendix:well-formedness}

The compiler has limitations on what expressions can be accelerated. These limitations also depend on which accelerate backend we use. First, we define a set of assumptions on accelerated code involving properties we need to check dynamically (Section~\ref{subsec:appendix:wf-runtime}). Second, we present the static well-formedness rules for the CUDA and Futhark backends (Section~\ref{subsec:appendix:wf-static}).

\subsection{Dynamic Assumptions}
\label{subsec:appendix:wf-runtime}

We define a set of assumptions that must hold at runtime in accelerated code.

We use $|x|$ to denote the length of a sequence $x$. In Futhark arrays must be \emph{regular}, meaning that for an array of arrays $[x_1, \ldots, x_n]$ we must have $|x_1| = \ldots = |x_n|$. Sequences in PMExpr, which we translate to arrays, do not have this restriction. Thus, we assume all sequences are regular in accelerated code targeting Futhark. Further, we assume constraints on sizes are annotated as required by Futhark.

The rank of a tensor is unknown at compile-time. In the CUDA backend, we represent their dimensions using a fixed-size array. We make the maximum rank of tensors configurable as in, for example, cuDNN\footnote{\url{https://docs.nvidia.com/deeplearning/cudnn/api/index.html\#cudnnSetTensorNdDescriptor}}. Our compiler uses $3$ as a default value to cover common use cases while also reducing wasted memory. We assume all tensors have dimensions no more than this maximum value. Our assumptions on regularity and tensor rank are checked at runtime for programs compiled in debug mode.

\begin{figure*}[t!]
\footnotesize
\begin{center}

\begin{prooftree}
\hypo{\Gamma' = \Gamma \cup \{(x, T)\}}
\hypo{\Gamma' \vdash_{EX} e : T'}
\infer2[(\textsc{WF-EX-Lam})]{\Gamma \vdash_{EX} \typedlamexpr{x}{T}{e} : T'}
\end{prooftree}
\hspace{5pt}
\begin{prooftree}
\hypo{\Gamma \vdash_{EX} e_1 : T}
\infer[no rule]1{\Gamma' \vdash_{EX} e_2 : T'}
\hypo{\Gamma \vdash_{PX} (e_1, p) : \Gamma'}
\infer[no rule]1{\Gamma \vdash_{EX} e_3 : T'}
\infer[no rule]2{T' \neq T_a \rightarrow T_b}
\infer1[(\textsc{WF-EX-Match})]{\Gamma \vdash_{EX} \matchexpr{e_1}{p}{e_2}{e_3} : T'}
\end{prooftree}
\hspace{5pt}
\begin{prooftree}
\infer0[(\textsc{WF-EX-Never})]{\Gamma \vdash_{EX} \texttt{never} : T}
\end{prooftree}
\hspace{5pt}
\begin{prooftree}
\hypo{\forall i \in 1 \ldots n.\ \Gamma \vdash_{EX} e_i : T_i}
\infer1[(\textsc{WF-EX-Record})]{\Gamma \vdash_{EX} \{l_i = e_i\}^{i \in 1 \ldots n} : \{l_i : T_i\}^{i \in 1 \ldots n}}
\end{prooftree}
\hspace{5pt}
\begin{prooftree}
\hypo{\forall i \in 1 \ldots n.\ \Gamma \vdash_{EX} e_i : T}
\infer1[(\textsc{WF-EX-Seq})]{\Gamma \vdash_{EX} [e_i]^{i \in 1 \ldots n}}
\end{prooftree}
\hspace{5pt}
\begin{prooftree}
\hypo{x : T \in \Gamma}
\infer1[(\textsc{WF-EX-Var})]{\Gamma \vdash_{EX} x : T}
\end{prooftree}
\hspace{5pt}
\begin{prooftree}
\hypo{c : T \in \Sigma_X}
\infer1[(\textsc{WF-EX-Const})]{\Gamma \vdash_{EX} c : T}
\end{prooftree}
\hspace{5pt}

\begin{prooftree}
\hypo{T \in T_g}
\infer1[(\textsc{WF-TX-Gr})]{\vdash_{TX} T}
\end{prooftree}
\hspace{5pt}
\begin{prooftree}
\hypo{\vdash_{TX} T_1}
\hypo{\vdash_{TX} T_2}
\infer2[(\textsc{WF-TX-Arrow})]{\vdash_{TX} T_1 \rightarrow T_2}
\end{prooftree}
\hspace{5pt}
\begin{prooftree}
\hypo{\forall i \in 1 \ldots n.\ \vdash_{TX} T_i \land T_i \neq T_a \rightarrow T_b}
\infer1[(\textsc{WF-TX-Rec})]{\vdash_{TX} \{l_i : T_i\}^{i \in 1 \ldots n}}
\end{prooftree}
\hspace{5pt}
\begin{prooftree}
\hypo{\vdash_{TX} T}
\hypo{T \neq T_a \rightarrow T_b}
\infer2[(\textsc{WF-TX-Seq})]{\vdash_{TX} [T]}
\end{prooftree}
\hspace{5pt}

\begin{prooftree}
\hypo{\Gamma' = \bigcup_{i \in 1 \ldots n} \Gamma_i}
\hypo{e_i = e.l_i}
\infer[no rule]2{\forall i \in 1 \ldots n.\ \Gamma \vdash_{PX} (e_i, p) : \Gamma_i}
\infer1[(\textsc{WF-PX-Rec})]{\Gamma \vdash_{PX} (e, \{l_i = p_i\}^{\forall i \in 1 \ldots n}) : \Gamma'}
\end{prooftree}
\hspace{5pt}
\begin{prooftree}
\hypo{\Gamma \vdash_{EX} e : T}
\hypo{\Gamma' = \Gamma \cup \{(x_p : T)\}}
\infer2[(\textsc{WF-PX-Var})]{\Gamma \vdash_{PX} (e, x_p) : \Gamma'}
\end{prooftree}
\hspace{5pt}
\begin{prooftree}
\hypo{c_p : T \in \Sigma_X}
\hypo{T \in T_g}
\infer2[(\textsc{WF-PX-Const})]{\Gamma \vdash_{PX} (e, c_p) : \Gamma}
\end{prooftree}
\end{center}
\caption{Definition of well-formedness rules shared by the CUDA and Futhark backends.}
\label{fig:wf-complete-common}
\end{figure*}

\begin{figure*}[t!]
\footnotesize
\begin{center}

\begin{prooftree}
\hypo{\blacktriangleright e_1 : T}
\infer[no rule]1{\Gamma' = \Gamma \cup \{(x, T)\}}
\hypo{\Gamma \vdash_{EC} e_1 : T}
\infer[no rule]1{\Gamma' \vdash_{EC} e_2 : T'}
\infer2[(\textsc{WF-EC-Let})]{\Gamma \vdash_{EC} \typedletexpr{x}{T}{e_1}{e_2} : T'}
\end{prooftree}
\hspace{5pt}
\begin{prooftree}
\hypo{\Gamma' = \Gamma \cup \left(\bigcup_{i \in 1 \ldots n} \{(x_i, T_i)\}\right)}
\hypo{\Gamma' \vdash_{EC} e : T}
\infer[no rule]2{\forall i \in 1 \ldots n.\ \Gamma' \vdash_{EC} e_i : T_i \land \blacktriangleright e_i : T_i}
\infer1[(\textsc{WF-EC-RecLet})]{\Gamma \vdash_{EC} \typedreclet{x}{T}{e}{e} : T}
\end{prooftree}
\hspace{5pt}
\begin{prooftree}
\hypo{\Gamma \vdash_{EC} e_1 : T_2 \rightarrow \ldots \rightarrow T_n \rightarrow T}
\hypo{T \neq T_a \rightarrow T_b}
\infer[no rule]2{\forall i \in 2 \ldots n.\ \Gamma \vdash_{EC} e_i : T_i}
\infer1[(\textsc{WF-EC-App})]{\Gamma \vdash_{EC} e_1\ e_2\ \ldots e_n : T}
\end{prooftree}
\hspace{5pt}
\begin{prooftree}
\hypo{\Gamma \vdash_{EC} e_1 : \msf{Int}}
\hypo{\Gamma \triangleright e_2 : \msf{Int} \rightarrow \{\}}
\infer2[(\textsc{WF-EC-Loop})]{\Gamma \vdash_{EC} \loopexpr{e_1}{e_2} : T}
\end{prooftree}
\hspace{5pt}

\begin{prooftree}
\hypo{\vdash_{TC} T}
\hypo{T = \msf{Int} \lor T = \msf{Float}}
\infer2[(\textsc{WF-TC-Ten})]{\vdash_{TC} \msf{Tensor}[T]}
\end{prooftree}

\begin{prooftree}
\hypo{T \neq T_a \rightarrow T_b}
\infer1[(\textsc{WF-BC-1})]{\blacktriangleright\ e : T}
\end{prooftree}
\hspace{5pt}
\begin{prooftree}
\hypo{\blacktriangleright\ e : T'}
\hypo{T \neq T_a \rightarrow T_b}
\infer2[(\textsc{WF-BC-2})]{\blacktriangleright\ \typedlamexpr{x}{T}{e} : T \rightarrow T'}
\end{prooftree}
\hspace{5pt}

\begin{prooftree}
\hypo{x : T \in \Gamma}
\infer1[(\textsc{WF-HC-Var})]{\Gamma\ \triangleright\ x : T}
\end{prooftree}
\hspace{5pt}
\begin{prooftree}
\hypo{\Gamma\ \triangleright\ e_1 : T_2 \rightarrow \ldots \rightarrow T_n \rightarrow T}
\infer[no rule]1{\forall i \in 2 \ldots n.\ \Gamma\ \triangleright\ e_i : T_i}
\infer1[(\textsc{WF-HC-App})]{\Gamma\ \triangleright\ e_1\ e_2 \ldots e_n : T}
\end{prooftree}
\hspace{5pt}

\end{center}
\caption{Definition of well-formedness rules specific to the CUDA backend.}
\label{fig:wf-complete-cuda}
\end{figure*}

\begin{figure*}[t!]
\footnotesize
\begin{center}
\begin{prooftree}
\hypo{\Gamma \vdash_{EF} e_1 : T}
\hypo{\Gamma' \vdash_{EF} e_2 : T'}
\infer[no rule]2{\Gamma' = \Gamma \cup \{(x, T)\}}
\infer1[(\textsc{WF-EF-Let})]{\Gamma \vdash_{EF} \typedletexpr{x}{T}{e_1}{e_2} : T'}
\end{prooftree}
\hspace{5pt}
\begin{prooftree}
\hypo{\Gamma \vdash_{EF} e_1 : T' \rightarrow T}
\hypo{\Gamma \vdash_{EF} e_2 : T'}
\infer2[(\textsc{WF-EF-App})]{\Gamma \vdash_{EF} e_1\ e_2 : T}
\end{prooftree}
\hspace{5pt}
\begin{prooftree}
\hypo{\Gamma \vdash_{EF} e_1 : T \rightarrow T'}
\hypo{\Gamma \vdash_{EF} e_2 : [T]}
\infer2[(\textsc{WF-EF-Map})]{\Gamma \vdash_{EF} \mapexpr{e_1}{e_2} : [T']}
\end{prooftree}
\hspace{5pt}
\begin{prooftree}
\hypo{\Gamma \vdash_{EF} e_2 : [T_1]}
\hypo{\Gamma \vdash_{EF} e_3 : [T_2]}
\infer[no rule]2{\Gamma \vdash_{EF} e_1 : T_1 \rightarrow T_2 \rightarrow T}
\infer1[(\textsc{WF-EF-Map2})]{\Gamma \vdash_{EF} \maptexpr{e_1}{e_2}{e_3} : [T]}
\end{prooftree}
\hspace{5pt}
\begin{prooftree}
\hypo{\Gamma \vdash_{EF} e_1 : T \rightarrow T \rightarrow T}
\infer[no rule]1{\Gamma \vdash_{EF} e_3 : [T]}
\hypo{\Gamma \vdash_{EF} e_2 : T}
\infer[no rule]1{T \neq T_a \rightarrow T_b}
\infer2[(\textsc{WF-EF-Red})]{\Gamma \vdash_{EF} \reduceexpr{e_1}{e_2}{e_3} : T}
\end{prooftree}
\hspace{5pt}
\begin{prooftree}
\hypo{\Gamma \vdash_{EF} e_1 : [[T]]}
\infer1[(\textsc{WF-EF-Flat})]{\Gamma \vdash_{EF} \flattenexpr{e_1} : [T]}
\end{prooftree}
\end{center}
\caption{Definition of well-formedness rules specific to the Futhark backend.}
\label{fig:wf-complete-futhark}
\end{figure*}

We also make assumptions about the parallel expressions. In particular, they must yield the same result regardless of execution order. While there are approaches to verify assumptions of deterministic parallelism statically~\cite{BocchinoADAHKOSSV09,KuperTKN14,HallerGES16,HelmKKHESM20}, they either require complex type systems or do not support our input language. Moreover, there are no efficient methods for verifying these assumptions dynamically. Therefore, the user is responsible for verifying them. For example, we assume for a reduce expression \mcoreinline{reduce f 0 s} that \mcoreinline{f} is associative and \mcoreinline{0} is the neutral element of \mcoreinline{f}.

\subsection{Static Rules}
\label{subsec:appendix:wf-static}

We define the static well-formedness rules for the CUDA and Futhark backends. Due to the similarities of the backends, we define a set of rules shared by both backends in Figure~\ref{fig:wf-complete-common}. We define rules specific to the CUDA backend in Figure~\ref{fig:wf-complete-cuda} and the rules for the Futhark backend in Figure~\ref{fig:wf-complete-futhark}.

We define rules over expressions ($E$), types ($T$), and patterns ($P$), which are pairwise combined with the two backends CUDA ($C$) and Futhark ($F$). That is, there are six different relations marked with $\mathit{EC}$, $\mathit{TC}$, $\mathit{PC}$, $\mathit{EF}$, $\mathit{TF}$, and $\mathit{PF}$. For instance, the relation for expressions in the CUDA backend is marked with $EC$. In this case, we say that an expression $e$ with well-formed type $T$ is well-formed in an environment $\Gamma$ iff $\Gamma \vdash_{EC} e : T$, where $\Gamma$ denotes an environment of pairs of identifiers and types, $(x, T)$. We do not need the environment $\Gamma$ for the well-formedness of types, as the language has no type variables. A type is well-formed in backend $X$ iff $\vdash_{TX} T$.

In several cases, the rules for both CUDA and Futhark are the same. Instead of repeating identical rules for both backends, we mark it with an $X$ to indicate that it implicitly represents two rules. For instance, a rule $\Gamma \vdash_{EX} e : T$ represents two rules, both for $EC$ and $EF$.

We discuss select rules to give an intuition of how they work and what we use them for. Consider the rule for match-expressions (\textsc{WF-EX-Match}) at the top center of Figure~\ref{fig:wf-complete-common}. We may bind variables in the pattern $p$ to the corresponding subexpressions of $e_1$. Thus, we need an updated environment $\Gamma'$ in the then-branch ($e_2$) containing these bound variables. Therefore, we define the well-formedness of a pattern $p$ matching on expression $e$ as well-formed in backend $X$ iff $\Gamma \vdash_{PX} (e, p) : \Gamma'$. Note also how the result of a match expression cannot be of a function type in either backend.

Consider the rules for let-expressions, \textsc{WF-EC-Let} and \textsc{WF-EF-Let}, at the top left of Figure~\ref{fig:wf-complete-cuda} and Figure~\ref{fig:wf-complete-futhark}, respectively. The rule for CUDA includes a premise using an auxiliary relation $\blacktriangleright e : T$, defined in \textsc{WF-BC-1} and \textsc{WF-BC-2}. This premise prevents user-defined functions from using or producing higher-order functions, as this is not supported.

The rule for loop expressions (\textsc{WF-EC-Loop}) uses a different auxiliary relation of the form $\Gamma\ \triangleright\ e : T$, used for higher-order functions (see the defining rules \textsc{WF-HC-Var} and \textsc{WF-HC-App} at the bottom of Figure~\ref{fig:wf-complete-cuda}). The rule for applications, \textsc{WF-HC-App}, is a relaxed variant of \textsc{WF-EC-App}, which allows the result to be a higher-order function. These rules are needed to enable the iteration function of a loop to use free variables (they are added as parameters to the iteration function by the lambda lifting). We assume ANF has been applied, but where applications are not lifted out of the higher-order function argument in parallel expressions. Therefore, the iteration function must be a variable or an application.

Assume that $e_C$ and $e_F$ represent the expressions resulting from the classification for the CUDA and Futhark backends. We say that the input program is well-formed if and only if, for well-formed types $T_1$ and $T_2$, it holds that
\[
\Gamma \vdash_{EC} e_C : T_1 \land \Gamma \vdash_{EF} e_F : T_2
\]

Assume that we have an input program for which the assumptions of Section~\ref{subsec:appendix:wf-runtime} are valid. If the program is well-typed and well-formed according to the well-formedness rules, our compiler ensures that all accelerated expressions are observationally equivalent to \mcoreinline{e}. Our implementation of the well-formedness checks is tested on a suite of example and benchmark programs. It is closely based on the definition of the well-formedness rules.

%% file: paper.bbl

\begin{thebibliography}{27}


\ifx \showCODEN    \undefined \def \showCODEN     #1{\unskip}     \fi
\ifx \showDOI      \undefined \def \showDOI       #1{#1}\fi
\ifx \showISBNx    \undefined \def \showISBNx     #1{\unskip}     \fi
\ifx \showISBNxiii \undefined \def \showISBNxiii  #1{\unskip}     \fi
\ifx \showISSN     \undefined \def \showISSN      #1{\unskip}     \fi
\ifx \showLCCN     \undefined \def \showLCCN      #1{\unskip}     \fi
\ifx \shownote     \undefined \def \shownote      #1{#1}          \fi
\ifx \showarticletitle \undefined \def \showarticletitle #1{#1}   \fi
\ifx \showURL      \undefined \def \showURL       {\relax}        \fi
\providecommand\bibfield[2]{#2}
\providecommand\bibinfo[2]{#2}
\providecommand\natexlab[1]{#1}
\providecommand\showeprint[2][]{arXiv:#2}

\bibitem[Bocchino~Jr et~al\mbox{.}(2009)]%
        {BocchinoADAHKOSSV09}
\bibfield{author}{\bibinfo{person}{Robert~L Bocchino~Jr}, \bibinfo{person}{Vikram~S Adve}, \bibinfo{person}{Danny Dig}, \bibinfo{person}{Sarita~V Adve}, \bibinfo{person}{Stephen Heumann}, \bibinfo{person}{Rakesh Komuravelli}, \bibinfo{person}{Jeffrey Overbey}, \bibinfo{person}{Patrick Simmons}, \bibinfo{person}{Hyojin Sung}, {and} \bibinfo{person}{Mohsen Vakilian}.} \bibinfo{year}{2009}\natexlab{}.
\newblock \showarticletitle{A type and effect system for deterministic parallel Java}. In \bibinfo{booktitle}{\emph{Proceedings of the 24th Annual {ACM} {SIGPLAN} Conference on Object-Oriented Programming, Systems, Languages, and Applications, {OOPSLA} 2009, October 25-29, 2009, Orlando, Florida, {USA}}}. \bibinfo{pages}{97--116}.
\newblock
\urldef\tempurl%
\url{https://doi.org/10.1145/1640089.1640097}
\showDOI{\tempurl}


\bibitem[Broman(2019)]%
        {broman2019vision}
\bibfield{author}{\bibinfo{person}{David Broman}.} \bibinfo{year}{2019}\natexlab{}.
\newblock \showarticletitle{A Vision of Miking: Interactive Programmatic Modeling, Sound Language Composition, and Self-Learning Compilation}. In \bibinfo{booktitle}{\emph{Proceedings of the 12th {ACM} {SIGPLAN} International Conference on Software Language Engineering, {SLE} 2019, Athens, Greece, October 20-22, 2019}}, \bibfield{editor}{\bibinfo{person}{Oscar Nierstrasz}, \bibinfo{person}{Jeff Gray}, {and} \bibinfo{person}{Bruno~C. d.~S.~Oliveira}} (Eds.). \bibinfo{publisher}{{ACM}}, \bibinfo{pages}{55--60}.
\newblock


\bibitem[Catanzaro et~al\mbox{.}(2011)]%
        {catanzaro2011copperhead}
\bibfield{author}{\bibinfo{person}{Bryan Catanzaro}, \bibinfo{person}{Michael Garland}, {and} \bibinfo{person}{Kurt Keutzer}.} \bibinfo{year}{2011}\natexlab{}.
\newblock \showarticletitle{Copperhead: compiling an embedded data parallel language}. In \bibinfo{booktitle}{\emph{Proceedings of the 16th {ACM} {SIGPLAN} Symposium on Principles and Practice of Parallel Programming, {PPOPP} 2011, San Antonio, TX, USA, February 12-16, 2011}}, \bibfield{editor}{\bibinfo{person}{Calin Cascaval} {and} \bibinfo{person}{Pen{-}Chung Yew}} (Eds.). \bibinfo{publisher}{{ACM}}, \bibinfo{pages}{47--56}.
\newblock


\bibitem[Chakravarty et~al\mbox{.}(2011)]%
        {chakravarty2011accelerating}
\bibfield{author}{\bibinfo{person}{Manuel M.~T. Chakravarty}, \bibinfo{person}{Gabriele Keller}, \bibinfo{person}{Sean Lee}, \bibinfo{person}{Trevor~L. McDonell}, {and} \bibinfo{person}{Vinod Grover}.} \bibinfo{year}{2011}\natexlab{}.
\newblock \showarticletitle{Accelerating Haskell array codes with multicore GPUs}. In \bibinfo{booktitle}{\emph{Proceedings of the {POPL} 2011 Workshop on Declarative Aspects of Multicore Programming, {DAMP} 2011, Austin, TX, USA, January 23, 2011}}, \bibfield{editor}{\bibinfo{person}{Manuel Carro} {and} \bibinfo{person}{John~H. Reppy}} (Eds.). \bibinfo{publisher}{{ACM}}, \bibinfo{pages}{3--14}.
\newblock


\bibitem[Dubach et~al\mbox{.}(2012)]%
        {dubach2012compiling}
\bibfield{author}{\bibinfo{person}{Christophe Dubach}, \bibinfo{person}{Perry Cheng}, \bibinfo{person}{Rodric~M. Rabbah}, \bibinfo{person}{David~F. Bacon}, {and} \bibinfo{person}{Stephen~J. Fink}.} \bibinfo{year}{2012}\natexlab{}.
\newblock \showarticletitle{Compiling a high-level language for GPUs: (via language support for architectures and compilers)}. In \bibinfo{booktitle}{\emph{{ACM} {SIGPLAN} Conference on Programming Language Design and Implementation, {PLDI} '12, Beijing, China - June 11 - 16, 2012}}, \bibfield{editor}{\bibinfo{person}{Jan Vitek}, \bibinfo{person}{Haibo Lin}, {and} \bibinfo{person}{Frank Tip}} (Eds.). \bibinfo{publisher}{{ACM}}, \bibinfo{pages}{1--12}.
\newblock


\bibitem[Enmyren and Kessler(2010)]%
        {EnmurenEtAl:2010}
\bibfield{author}{\bibinfo{person}{Johan Enmyren} {and} \bibinfo{person}{Christoph~W Kessler}.} \bibinfo{year}{2010}\natexlab{}.
\newblock \showarticletitle{{SkePU: a multi-backend skeleton programming library for multi-GPU systems}}. In \bibinfo{booktitle}{\emph{Proceedings of the fourth international workshop on High-level parallel programming and applications}}. \bibinfo{pages}{5--14}.
\newblock


\bibitem[Ernstsson et~al\mbox{.}(2018)]%
        {ErnstssonEtal:2018}
\bibfield{author}{\bibinfo{person}{August Ernstsson}, \bibinfo{person}{Lu Li}, {and} \bibinfo{person}{Christoph Kessler}.} \bibinfo{year}{2018}\natexlab{}.
\newblock \showarticletitle{{SkePU 2: Flexible and type-safe skeleton programming for heterogeneous parallel systems}}.
\newblock \bibinfo{journal}{\emph{International Journal of Parallel Programming}} \bibinfo{volume}{46}, \bibinfo{number}{1} (\bibinfo{year}{2018}), \bibinfo{pages}{62--80}.
\newblock


\bibitem[Frostig et~al\mbox{.}(2018)]%
        {frostig2018compiling}
\bibfield{author}{\bibinfo{person}{Roy Frostig}, \bibinfo{person}{Matthew~James Johnson}, {and} \bibinfo{person}{Chris Leary}.} \bibinfo{year}{2018}\natexlab{}.
\newblock \showarticletitle{Compiling machine learning programs via high-level tracing}.
\newblock \bibinfo{journal}{\emph{Systems for Machine Learning}} \bibinfo{volume}{4}, \bibinfo{number}{9} (\bibinfo{year}{2018}).
\newblock


\bibitem[Hagedorn et~al\mbox{.}(2020)]%
        {hagedorn2020achieving}
\bibfield{author}{\bibinfo{person}{Bastian Hagedorn}, \bibinfo{person}{Johannes Lenfers}, \bibinfo{person}{Thomas Koehler}, \bibinfo{person}{Xueying Qin}, \bibinfo{person}{Sergei Gorlatch}, {and} \bibinfo{person}{Michel Steuwer}.} \bibinfo{year}{2020}\natexlab{}.
\newblock \showarticletitle{Achieving high-performance the functional way: a functional pearl on expressing high-performance optimizations as rewrite strategies}.
\newblock \bibinfo{journal}{\emph{Proc. {ACM} Program. Lang.}} \bibinfo{volume}{4}, \bibinfo{number}{{ICFP}} (\bibinfo{year}{2020}), \bibinfo{pages}{92:1--92:29}.
\newblock


\bibitem[Haller et~al\mbox{.}(2016)]%
        {HallerGES16}
\bibfield{author}{\bibinfo{person}{Philipp Haller}, \bibinfo{person}{Simon Geries}, \bibinfo{person}{Michael Eichberg}, {and} \bibinfo{person}{Guido Salvaneschi}.} \bibinfo{year}{2016}\natexlab{}.
\newblock \showarticletitle{Reactive Async: expressive deterministic concurrency}. In \bibinfo{booktitle}{\emph{Proceedings of the 7th {ACM} {SIGPLAN} Symposium on Scala, SCALA@SPLASH 2016, Amsterdam, Netherlands, October 30 - November 4, 2016}}. \bibinfo{pages}{11--20}.
\newblock
\urldef\tempurl%
\url{https://doi.org/10.1145/2998392.2998396}
\showDOI{\tempurl}


\bibitem[Helm et~al\mbox{.}(2020)]%
        {HelmKKHESM20}
\bibfield{author}{\bibinfo{person}{Dominik Helm}, \bibinfo{person}{Florian K{\"{u}}bler}, \bibinfo{person}{Jan~Thomas K{\"{o}}lzer}, \bibinfo{person}{Philipp Haller}, \bibinfo{person}{Michael Eichberg}, \bibinfo{person}{Guido Salvaneschi}, {and} \bibinfo{person}{Mira Mezini}.} \bibinfo{year}{2020}\natexlab{}.
\newblock \showarticletitle{A programming model for semi-implicit parallelization of static analyses}. In \bibinfo{booktitle}{\emph{{ISSTA} '20: 29th {ACM} {SIGSOFT} International Symposium on Software Testing and Analysis, Virtual Event, USA, July 18-22, 2020}}. \bibinfo{pages}{428--439}.
\newblock
\urldef\tempurl%
\url{https://doi.org/10.1145/3395363.3397367}
\showDOI{\tempurl}


\bibitem[Henriksen et~al\mbox{.}(2017)]%
        {henriksen2017futhark}
\bibfield{author}{\bibinfo{person}{Troels Henriksen}, \bibinfo{person}{Niels G.~W. Serup}, \bibinfo{person}{Martin Elsman}, \bibinfo{person}{Fritz Henglein}, {and} \bibinfo{person}{Cosmin~E. Oancea}.} \bibinfo{year}{2017}\natexlab{}.
\newblock \showarticletitle{Futhark: purely functional GPU-programming with nested parallelism and in-place array updates}. In \bibinfo{booktitle}{\emph{Proceedings of the 38th {ACM} {SIGPLAN} Conference on Programming Language Design and Implementation, {PLDI} 2017, Barcelona, Spain, June 18-23, 2017}}, \bibfield{editor}{\bibinfo{person}{Albert Cohen} {and} \bibinfo{person}{Martin~T. Vechev}} (Eds.). \bibinfo{publisher}{{ACM}}, \bibinfo{pages}{556--571}.
\newblock


\bibitem[Ishizaki et~al\mbox{.}(2015)]%
        {ishizaki2015compiling}
\bibfield{author}{\bibinfo{person}{Kazuaki Ishizaki}, \bibinfo{person}{Akihiro Hayashi}, \bibinfo{person}{Gita Koblents}, {and} \bibinfo{person}{Vivek Sarkar}.} \bibinfo{year}{2015}\natexlab{}.
\newblock \showarticletitle{Compiling and Optimizing Java 8 Programs for {GPU} Execution}. In \bibinfo{booktitle}{\emph{2015 International Conference on Parallel Architectures and Compilation, {PACT} 2015, San Francisco, CA, USA, October 18-21, 2015}}. \bibinfo{publisher}{{IEEE} Computer Society}, \bibinfo{pages}{419--431}.
\newblock


\bibitem[Johnsson(1985)]%
        {johnsson1985lambda}
\bibfield{author}{\bibinfo{person}{Thomas Johnsson}.} \bibinfo{year}{1985}\natexlab{}.
\newblock \showarticletitle{Lambda Lifting: Transforming Programs to Recursive Equations}. In \bibinfo{booktitle}{\emph{Functional Programming Languages and Computer Architecture, {FPCA} 1985, Nancy, France, September 16-19, 1985, Proceedings}} \emph{(\bibinfo{series}{Lecture Notes in Computer Science}, Vol.~\bibinfo{volume}{201})}, \bibfield{editor}{\bibinfo{person}{Jean{-}Pierre Jouannaud}} (Ed.). \bibinfo{publisher}{Springer}, \bibinfo{pages}{190--203}.
\newblock


\bibitem[Kuper et~al\mbox{.}(2014)]%
        {KuperTKN14}
\bibfield{author}{\bibinfo{person}{Lindsey Kuper}, \bibinfo{person}{Aaron Turon}, \bibinfo{person}{Neelakantan~R. Krishnaswami}, {and} \bibinfo{person}{Ryan~R. Newton}.} \bibinfo{year}{2014}\natexlab{}.
\newblock \showarticletitle{Freeze after writing: quasi-deterministic parallel programming with LVars}. In \bibinfo{booktitle}{\emph{The 41st Annual {ACM} {SIGPLAN-SIGACT} Symposium on Principles of Programming Languages, {POPL} '14, San Diego, CA, USA, January 20-21, 2014}}. \bibinfo{pages}{257--270}.
\newblock
\urldef\tempurl%
\url{https://doi.org/10.1145/2535838.2535842}
\showDOI{\tempurl}


\bibitem[Lam et~al\mbox{.}(2015)]%
        {lam2015numba}
\bibfield{author}{\bibinfo{person}{Siu~Kwan Lam}, \bibinfo{person}{Antoine Pitrou}, {and} \bibinfo{person}{Stanley Seibert}.} \bibinfo{year}{2015}\natexlab{}.
\newblock \showarticletitle{Numba: a LLVM-based Python {JIT} compiler}. In \bibinfo{booktitle}{\emph{Proceedings of the Second Workshop on the {LLVM} Compiler Infrastructure in HPC, {LLVM} 2015, Austin, Texas, USA, November 15, 2015}}, \bibfield{editor}{\bibinfo{person}{Hal Finkel}} (Ed.). \bibinfo{publisher}{{ACM}}, \bibinfo{pages}{7:1--7:6}.
\newblock


\bibitem[Lei{\ss}a et~al\mbox{.}(2018)]%
        {leissa2018anydsl}
\bibfield{author}{\bibinfo{person}{Roland Lei{\ss}a}, \bibinfo{person}{Klaas Boesche}, \bibinfo{person}{Sebastian Hack}, \bibinfo{person}{Ars{\`{e}}ne P{\'{e}}rard{-}Gayot}, \bibinfo{person}{Richard Membarth}, \bibinfo{person}{Philipp Slusallek}, \bibinfo{person}{Andr{\'{e}} M{\"{u}}ller}, {and} \bibinfo{person}{Bertil Schmidt}.} \bibinfo{year}{2018}\natexlab{}.
\newblock \showarticletitle{AnyDSL: a partial evaluation framework for programming high-performance libraries}.
\newblock \bibinfo{journal}{\emph{Proc. {ACM} Program. Lang.}} \bibinfo{volume}{2}, \bibinfo{number}{{OOPSLA}} (\bibinfo{year}{2018}), \bibinfo{pages}{119:1--119:30}.
\newblock


\bibitem[OpenMP Architecture Review Board(2021)]%
        {openmp}
OpenMP Architecture Review Board \bibinfo{year}{2021}\natexlab{}.
\newblock \bibinfo{booktitle}{\emph{OpenMP Application Programming Interface}}.
\newblock OpenMP Architecture Review Board.
\newblock


\bibitem[Ragan{-}Kelley et~al\mbox{.}(2013)]%
        {kelley2013halide}
\bibfield{author}{\bibinfo{person}{Jonathan Ragan{-}Kelley}, \bibinfo{person}{Connelly Barnes}, \bibinfo{person}{Andrew Adams}, \bibinfo{person}{Sylvain Paris}, \bibinfo{person}{Fr{\'{e}}do Durand}, {and} \bibinfo{person}{Saman~P. Amarasinghe}.} \bibinfo{year}{2013}\natexlab{}.
\newblock \showarticletitle{Halide: a language and compiler for optimizing parallelism, locality, and recomputation in image processing pipelines}. In \bibinfo{booktitle}{\emph{{ACM} {SIGPLAN} Conference on Programming Language Design and Implementation, {PLDI} '13, Seattle, WA, USA, June 16-19, 2013}}, \bibfield{editor}{\bibinfo{person}{Hans{-}Juergen Boehm} {and} \bibinfo{person}{Cormac Flanagan}} (Eds.). \bibinfo{publisher}{{ACM}}, \bibinfo{pages}{519--530}.
\newblock


\bibitem[Rossbach et~al\mbox{.}(2013)]%
        {rossbach2013dandelion}
\bibfield{author}{\bibinfo{person}{Christopher~J. Rossbach}, \bibinfo{person}{Yuan Yu}, \bibinfo{person}{Jon Currey}, \bibinfo{person}{Jean{-}Philippe Martin}, {and} \bibinfo{person}{Dennis Fetterly}.} \bibinfo{year}{2013}\natexlab{}.
\newblock \showarticletitle{Dandelion: a compiler and runtime for heterogeneous systems}. In \bibinfo{booktitle}{\emph{{ACM} {SIGOPS} 24th Symposium on Operating Systems Principles, {SOSP} '13, Farmington, PA, USA, November 3-6, 2013}}, \bibfield{editor}{\bibinfo{person}{Michael Kaminsky} {and} \bibinfo{person}{Mike Dahlin}} (Eds.). \bibinfo{publisher}{{ACM}}, \bibinfo{pages}{49--68}.
\newblock


\bibitem[Steuwer et~al\mbox{.}(2017)]%
        {steuwer2017lift}
\bibfield{author}{\bibinfo{person}{Michel Steuwer}, \bibinfo{person}{Toomas Remmelg}, {and} \bibinfo{person}{Christophe Dubach}.} \bibinfo{year}{2017}\natexlab{}.
\newblock \showarticletitle{Lift: a functional data-parallel {IR} for high-performance {GPU} code generation}. In \bibinfo{booktitle}{\emph{Proceedings of the 2017 International Symposium on Code Generation and Optimization, {CGO} 2017, Austin, TX, USA, February 4-8, 2017}}, \bibfield{editor}{\bibinfo{person}{Vijay~Janapa Reddi}, \bibinfo{person}{Aaron Smith}, {and} \bibinfo{person}{Lingjia Tang}} (Eds.). \bibinfo{publisher}{{ACM}}, \bibinfo{pages}{74--85}.
\newblock
\urldef\tempurl%
\url{http://dl.acm.org/citation.cfm?id=3049841}
\showURL{%
\tempurl}


\bibitem[Sujeeth et~al\mbox{.}(2014)]%
        {sujeeth2014delite}
\bibfield{author}{\bibinfo{person}{Arvind~K. Sujeeth}, \bibinfo{person}{Kevin~J. Brown}, \bibinfo{person}{HyoukJoong Lee}, \bibinfo{person}{Tiark Rompf}, \bibinfo{person}{Hassan Chafi}, \bibinfo{person}{Martin Odersky}, {and} \bibinfo{person}{Kunle Olukotun}.} \bibinfo{year}{2014}\natexlab{}.
\newblock \showarticletitle{Delite: {A} Compiler Architecture for Performance-Oriented Embedded Domain-Specific Languages}.
\newblock \bibinfo{journal}{\emph{{ACM} Trans. Embed. Comput. Syst.}} \bibinfo{volume}{13}, \bibinfo{number}{4s} (\bibinfo{year}{2014}), \bibinfo{pages}{134:1--134:25}.
\newblock


\bibitem[Svensson et~al\mbox{.}(2008)]%
        {svensson2008obsidian}
\bibfield{author}{\bibinfo{person}{Joel Svensson}, \bibinfo{person}{Mary Sheeran}, {and} \bibinfo{person}{Koen Claessen}.} \bibinfo{year}{2008}\natexlab{}.
\newblock \showarticletitle{Obsidian: {A} Domain Specific Embedded Language for Parallel Programming of Graphics Processors}. In \bibinfo{booktitle}{\emph{Implementation and Application of Functional Languages - 20th International Symposium, {IFL} 2008, Hatfield, UK, September 10-12, 2008. Revised Selected Papers}} \emph{(\bibinfo{series}{Lecture Notes in Computer Science}, Vol.~\bibinfo{volume}{5836})}, \bibfield{editor}{\bibinfo{person}{Sven{-}Bodo Scholz} {and} \bibinfo{person}{Olaf Chitil}} (Eds.). \bibinfo{publisher}{Springer}, \bibinfo{pages}{156--173}.
\newblock


\bibitem[Team et~al\mbox{.}(2016)]%
        {team2016theano}
\bibfield{author}{\bibinfo{person}{The Theano~Development Team}, \bibinfo{person}{Rami Al-Rfou}, \bibinfo{person}{Guillaume Alain}, \bibinfo{person}{Amjad Almahairi}, \bibinfo{person}{Christof Angermueller}, \bibinfo{person}{Dzmitry Bahdanau}, \bibinfo{person}{Nicolas Ballas}, \bibinfo{person}{Fr{\'e}d{\'e}ric Bastien}, \bibinfo{person}{Justin Bayer}, \bibinfo{person}{Anatoly Belikov}, {et~al\mbox{.}}} \bibinfo{year}{2016}\natexlab{}.
\newblock \showarticletitle{Theano: A Python framework for fast computation of mathematical expressions}.
\newblock \bibinfo{journal}{\emph{arXiv preprint arXiv:1605.02688}} (\bibinfo{year}{2016}).
\newblock


\bibitem[The OpenACC organization(2021)]%
        {openacc}
The OpenACC organization \bibinfo{year}{2021}\natexlab{}.
\newblock \bibinfo{booktitle}{\emph{The OpenACC Application Programming Interface}}.
\newblock The OpenACC organization.
\newblock


\bibitem[Tillet et~al\mbox{.}(2019)]%
        {tillet2019triton}
\bibfield{author}{\bibinfo{person}{Philippe Tillet}, \bibinfo{person}{Hsiang-Tsung Kung}, {and} \bibinfo{person}{David Cox}.} \bibinfo{year}{2019}\natexlab{}.
\newblock \showarticletitle{Triton: an intermediate language and compiler for tiled neural network computations}. In \bibinfo{booktitle}{\emph{Proceedings of the 3rd ACM SIGPLAN International Workshop on Machine Learning and Programming Languages}}. \bibinfo{pages}{10--19}.
\newblock


\bibitem[Zhou et~al\mbox{.}(2024)]%
        {zhou2024appy}
\bibfield{author}{\bibinfo{person}{Tong Zhou}, \bibinfo{person}{Jun Shirako}, {and} \bibinfo{person}{Vivek Sarkar}.} \bibinfo{year}{2024}\natexlab{}.
\newblock \showarticletitle{APPy: Annotated Parallelism for Python on GPUs}. In \bibinfo{booktitle}{\emph{Proceedings of the 33rd ACM SIGPLAN International Conference on Compiler Construction}}. \bibinfo{pages}{113--125}.
\newblock


\end{thebibliography}
